\documentclass[%
reprint,
groupedaddress,
amsmath,amssymb,
aps,
prx,
nolongbibliography, 
]{revtex4-2}

\setcitestyle{maxcitenames=3,url=false,doi=false,eprint=false}
\usepackage{graphicx}
\usepackage{dcolumn}
\usepackage{bm}
\usepackage{xr}
\usepackage{cancel}
\usepackage[markup=underlined]{changes} 
\usepackage{booktabs}
\usepackage{nicefrac}

\usepackage[colorlinks=true,
            linkcolor=blue,
            citecolor=blue,
            urlcolor=blue]{hyperref}

\usepackage{lipsum}

\usepackage{color}

\definecolor{waldgruen}{RGB}{34,139,34}

\definecolor{old}{RGB}{255,165,0}

\usepackage{tocbasic}
\DeclareNewTOC[%
  type=Equation,
  float
]{loe}

\usepackage{caption}
\usepackage{chngcntr}

\usepackage{ragged2e}

\makeatletter
\long\def\@makecaption#1#2{%
  \par
  \begingroup
    \small
    \justifying
    #1. #2\par
  \endgroup
}
\makeatother

\begin{document}
	\renewcommand*{\arraystretch}{1.1}

	\title{Tunable Signal Penetration and Response Plateaus in Bistable Mechanical Media}

	\author{Sven Pattloch}
	\affiliation{Applied Theoretical Physics - Computational Physics, Physikalisches Institut, Albert-Ludwigs-Universit\"at Freiburg, D-79104 Freiburg, Germany}
	\affiliation{Cluster of Excellence livMatS @ FIT - Freiburg Center for Interactive Materials and Bioinspired Technologies, Albert-Ludwigs-Universit\"at Freiburg, D-79110 Freiburg, Germany}
   	\author{Joachim Dzubiella}
	\affiliation{Applied Theoretical Physics - Computational Physics, Physikalisches Institut, Albert-Ludwigs-Universit\"at Freiburg, D-79104 Freiburg, Germany}
    \affiliation{Cluster of Excellence livMatS @ FIT - Freiburg Center for Interactive Materials and Bioinspired Technologies, Albert-Ludwigs-Universit\"at Freiburg, D-79110 Freiburg, Germany}
	
	\date{\today}

\begin{abstract}
The ability to dynamically process mechanical signals is crucial for advanced functionalities in soft robotics and mechanosensing, where classical viscoelastic materials often lack intrinsic tunability and quantitative design rules are missing. 
In this work, we show that incorporating internal bistability provides a simple but powerful mechanism to tailor response and signal attenuation in mechanical (meta)materials. 
In our model, the bistable elements switch discretely with a predefined intrinsic timescale between states distinguished by a change in potential energy $\epsilon$, equilibrium length $\Delta l$, and spring constant $\Delta k$.
The model is simulated numerically by microscopic Brownian dynamics coupled to discrete Poisson switching with rate $\nu$.  In addition we develop a nonlinear continuum (field) theory that describes the macroscopic strain relaxation.  Crucially, the model yields closed-form analytical solutions for the linear response and the spatial penetration depth, revealing two complementary phenomena: a universal screening mechanism (akin to the electrostatic `skin effect') that strongly reduces spatial signal penetration when the driving frequency exceeds the internal relaxation rate, and a frequency-insensitive response plateau emerging from timescale separation. The attenuation strength, expressed by a screening length, is controlled primarily by the conformational length change $\Delta l$, while both the attenuation regime and plateau position are tuneable via the switching rate $\nu$.
A systematic parameter study exposes a fundamental design trade-off: while larger conformational changes $\Delta l$ strengthen dissipation, they simultaneously raise the (elastic) energy barrier for state transitions, eventually triggering state-locking where the damping vanishes. Optimal signal attenuation thus emerges from a compromise between pronounced bistability and a surmountable barrier. Due to its analytical tractability and generic formulation, our framework provides explicit design rules for fine-tuning the adaptive response of mechanical bistable media. It is readily applicable to a wide range of experimental bistable systems -- from biopolymers to synthetic catch bonds and metamaterials -- providing a quantitative foundation for the predictive design of soft materials with frequency-selective mechanical response.
\end{abstract}

\maketitle

\section{Introduction}
The design of materials with adaptive functional properties is a central pursuit in modern condensed matter physics and engineering. By shifting focus from passive elasticity to systems operating far from equilibrium, the inclusion of internal degrees of freedom allows for the emergence of complex behaviors such as memory formation, self-organization, and time-dependent learning \cite{Mungan_2025b, Nagel_2023, Deutsch_2004}.
A particularly powerful and widely studied mechanism to achieve this functionality is bistability, where individual elements can exist in two distinct stable configurations. Phenomenologically, collections of such bistable units are the building blocks of hysteretic systems described by Preisach models and are crucial for the development of smart materials and metamaterials capable of storing processing history \cite{Keim_2019, Keim_2020, Paulsen_2025a, Paulsen_2025b, Keim_2019, Lindeman_2023, Lindeman_2025}. These capabilities are especially sought after in applications ranging from energy absorption and tunable damping in vitrimers \cite{Zhao_2025, Tanaka_2024, Ast_2009} to soft robotics and compliant actuation, where the integration of signal processing, energy dissipation, and tunable stiffness directly at the material level could revolutionize human-robot interaction and locomotion strategies \cite{Vanderborght_2008, Chen_2018, Laffranchi_2013, Culmer_2010, Pratt_1995}.

The interplay between mechanical deformation and internal state transitions has been explored across various scales and architectures. In polymer physics and biophysics, bistability serves as a fundamental model for the force-extension response of biopolymers and the unfolding of repeat proteins, where conformational transitions dictate the large-scale mechanical response \cite{Rief_1998, Giordano_2017, Benedito_2020, Makarov_2009, Matevosyan_2026}. Similarly, transient polymer networks and reversibly crosslinked polymer melts exhibit tunable damping, self-healing, and anomalous tension propagation derived from reversible bond kinetics \cite{Karmakar_2025, Saito_2015, Tanaka_2024, Zhao_2025, Tian_2022}. In the realm of mechanical metamaterials, geometric bistability—realized through structural buckling, origami-inspired designs, or shape-memory elements—has enabled the creation of tunable architectures capable of sequential deformation, strain stiffening, auxeticity, and programmable multistability \cite{Rafsanjani_2016, Nitecki_2021, Kamrava_2017, Zhang_2021, Che_2016, Raney_2016, Ghavidelnia_2023, Roller_2024}. These internal dynamics serve as the physical basis for mechanical memory, allowing materials to retain a trace of past driving histories, such as return point memory, and even facilitating mechanical training, allosteric-inspired responses, and epistatic pathways \cite{Mungan_2022, Mandal_2024, Mandal_2025, Rocks_2017, Alqatari_2024}. While current research heavily emphasizes non-reciprocal transport enabled by active or energetically asymmetric driving \cite{Veenstra_2024, Brauns_2024, Librandi_2021, Scheibner_2020, Brandenbourger_2019} and mechanochemical feedback loops \cite{Sarkar_2025, Perez_Verdugo_2024, Goulefack_2025, Geher_Herczegh_2024, Watabe_2024}, which have been shown to fundamentally control energy dissipation and entropy production in active solids \cite{Cocconi_2025}, the intrinsic dissipative behavior arising from purely bistable kinetics—such as the emergence of a mechanical skin effect or plastic screening in amorphous solids—remains a complex frontier \cite{Kumar_2024, Lemaitre_2021, Sipling_2025, Biot_1966, Tarjus_2026}.

The interplay between mechanical deformation and internal state transitions has been explored across various scales and architectures. In polymer physics and biophysics, bistability serves as a fundamental model for the force-extension response of biopolymers and the unfolding of repeat proteins, where conformational transitions dictate the large-scale mechanical response \cite{Rief_1998, Giordano_2017, Benedito_2020, Makarov_2009, Matevosyan_2026}. Similarly, transient polymer networks and reversibly crosslinked polymer melts exhibit tunable damping, self-healing, and anomalous tension propagation derived from reversible bond kinetics \cite{Karmakar_2025, Saito_2015, Tanaka_2024, Zhao_2025, Tian_2022}. In the realm of mechanical metamaterials, geometric bistability—realized through structural buckling, origami-inspired designs, or shape-memory elements—has enabled the creation of tunable architectures capable of sequential deformation, strain stiffening, auxeticity, and multistability \cite{Rafsanjani_2016, Nitecki_2021, Kamrava_2017, Zhang_2021, Che_2016, Raney_2016, Ghavidelnia_2023, Roller_2024}. These internal dynamics serve as the physical basis for mechanical memory, allowing materials to retain a trace of past driving histories, such as return point memory, and even facilitating mechanical training, allosteric-inspired responses, and epistatic pathways \cite{Mungan_2022, Mandal_2024, Mandal_2025, Rocks_2017, Alqatari_2024}.

While current research heavily emphasizes non-reciprocal transport enabled by active or energetically asymmetric driving \cite{Veenstra_2024, Brauns_2024, Librandi_2021, Scheibner_2020, Brandenbourger_2019} and mechanochemical feedback loops \cite{Sarkar_2025, Perez_Verdugo_2024, Goulefack_2025, Geher_Herczegh_2024, Watabe_2024}, which have been shown to fundamentally control energy dissipation and entropy production in active solids \cite{Cocconi_2025}, the intrinsic dissipative behavior arising from purely bistable kinetics—such as the emergence of a mechanical skin effect or plastic screening in amorphous solids—remains a complex frontier \cite{Kumar_2024, Lemaitre_2021, Sipling_2025, Biot_1966, Tarjus_2026}.

The foundation for modeling bistable lattice systems was historically established by the Krumhansl-Schrieffer Hamiltonian for structural phase transitions \cite{Krumhansl_1975}, describing the motion of domain walls and the stochastic escape processes characteristic of two-level systems \cite{Wada_1978, Karakurt_2007}. Extensive work has been done on the rate-dependent response of bistable chains, exhibiting hysteresis and force-extension plateaus \cite{Benichou_2016, Majumdar_2018}, noise-enhanced stability phenomena in fluctuating potentials \cite{Spagnolo_2004, Spagnolo_2004b}, and resonant activation under periodic drifts \cite{Herrmann_2005}. These systems constitute the microscopic realization of mechanical hysterons, whose collective behavior—ranging from interacting elements to catastrophic transitions—has been analyzed in depth to understand memory states, return point memory, and multi-bit storage in frustrated systems \cite{Muhaxheri_2024, Muhaxheri_2025, Teunisse_2025, Baconnier_2025, Chaudhuri_2008, Kimizuka_2010, Sevilla_2014, Dossetti_2015, Kedia_2023}. Moreover, advancements in experiments and fabrication have demonstrated the robustness of these principles in diverse materials, utilizing hydrogel elements with pH-switchable stiffening, embedded permanent magnets, and magnetorheological tunability \cite{Skarsetz_2022, Skarsetz_2024, Caballero-Russi_2025, Slesarenko_2020, Wenz_2023, Case_2013}. This includes the realization of controlled transition waves, propagating solitons, guided transition pathways with honeycomb geometries, and harnessing bistability for directional propulsion in untethered soft robots \cite{Jin_2020, Jin_2025, Liu_2024, Liu_2024b, Gutierrezprieto_2025, Tang_2025, Chen_2018}. Furthermore, recent progress in creating chemically fueled bistable crystals and chemomechanical coupling extends these concepts to molecular information processing and OR-port realization \cite{Schnitter_2022}. Recent demonstrations of coupled learning in elastic networks and reversible training in flow networks further highlight the potential of these systems for computational tasks \cite{Altman_2024, Altman_2025, Berry_2022}.

Bridging the gap between the atomistic kinetics of individual switching events and the macroscopic, spatiotemporal response of a driven chain poses a significant theoretical challenge. While exact analytical solutions exist for heterogeneous overdamped chains with harmonic interactions \cite{Saha_2026}, and dissipative effects in bistable chains have been studied in the rate-dependent limit with domain wall dynamics explored numerically \cite{Benichou_2016, Kimizuka_2010}, a coarse-grained theory linking the microscopic transition rate to the propagation and attenuation of mechanical waves in the overdamped, continuous limit is lacking. Specifically, it remains unclear how the intrinsic switching timescale interacts with the diffusive timescale of mechanical transport to produce \emph{two complementary phenomena}: a frequency-selective screening of mechanical signals (analogous to the `skin effect' in electromagnetism) \emph{and} a frequency-insensitive response plateau arising from timescale separation. In this work, we address this gap by establishing a continuum framework that maps microscopic switching rates and barrier-crossing kinetics onto macroscopic signal dissipation. By identifying how internal switching timescales dictate spatial penetration depths and compliance plateaus, our work provides concrete design principles for frequency-dependent signal attenuation. Beyond passive filtering, this quantitative understanding of internal relaxation timescales establishes an essential theoretical foundation required for future investigations into dynamic memory formation and mechanical training in driven bistable media.

\quad
\newline
\section{Model and Simulations}
\begin{figure}[h]
	\centering
		\includegraphics[width=\columnwidth]{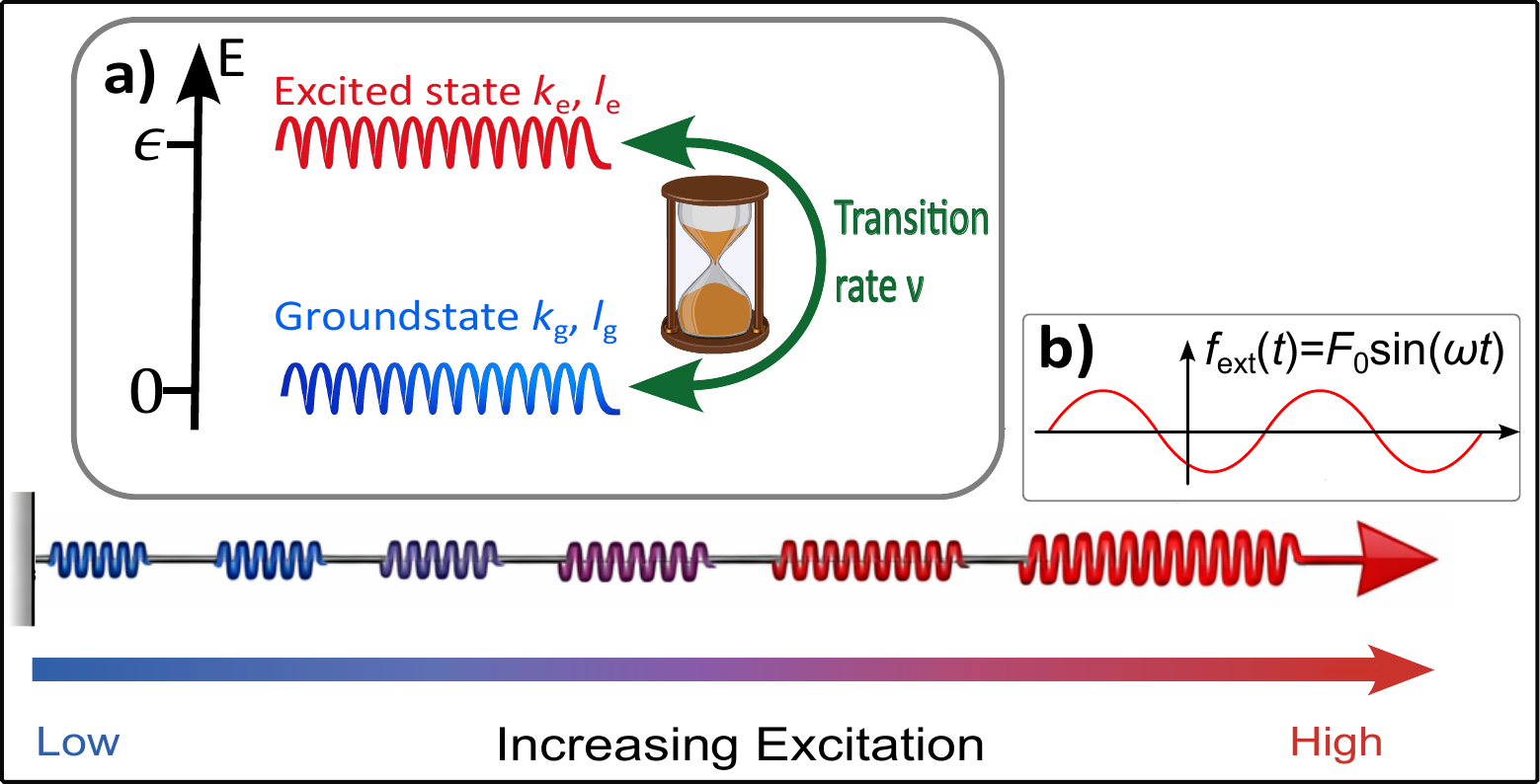}
	\caption{\textbf{Model sketch.} A one-dimensional chain of bistable springs is clamped at the left end (fixed boundary) and driven at the right end by a periodic external force $f_\mathrm{ext}(t)$ (inset \textbf{b})). The local deformation and excitation increases towards the right boundary due to spatial attenuation (damping) of the incoming signal. Inset \textbf{a}): bistable energy of a single spring, which can switch between a ground state (blue spring, with parameters $l_\mathrm{g}$, $k_\mathrm{g}$, energy zero) and an excited state (red spring, with parameters $l_\mathrm{e}$, $k_\mathrm{e}$, energy $\epsilon$). The switching occurs on the characteristic transition rate $\nu$.}
\label{fig:Model}
\end{figure}
As in our previous work on equilibrium only \cite{Pattloch_2025}, we consider a one-dimensional (1D) system of $N$ bistable springs connecting $N+1$ `particles' or `modules'. The model setup, sketched in Fig.~\ref{fig:Model}, consists of a 1D chain clamped at the left end (module position $x_0=0$ connected to first spring $i=1$) and driven at the free right end by an external force $f_{\rm ext}(t)$ acting on module $N$. Each spring can exist in two discrete states with equilibrium length $l_\mathrm{g}$ and spring constant $k_\mathrm{g}$ or the excited length $l_\mathrm{e} = l_\mathrm{g} + \Delta l$ and excited spring constant $k_\mathrm{e} = k_\mathrm{g} + \Delta k$, respectively. The potential energies of the ground and excited states differ by the energy $\epsilon$.

The Hamiltonian of the system thus reads
\begin{equation}
	\label{eq:Htot}
		\begin{split}
			{\cal H_\mathrm{tot}}
			&= \sum\limits_{j=1}^N\epsilon n_j + \sum\limits_{j=1}^N \frac{k_j}{2} \left(x_{j} - x_{j-1} - l_j \right)^2  - f_\mathrm{ext} x_N,
		\end{split}
\end{equation}
where $n_j = 0,1$ denotes the state of spring $j$, and $k_j = k_\mathrm{g}+n_j\Delta k$ and $l_j = l_\mathrm{g}+n_j\Delta l$.  Let $q$ and $p = 1-q$ denote the fraction of modules in the excited and ground state, respectively. In the quasistatic limit (cf.~\cite{Pattloch_2025}), the system minimizes the free energy, leading to the force-dependent equilibrium excitation probability
\begin{equation}
	\label{eq:qeq}
	q_\mathrm{eq}\left(f\right) = \frac{\chi}{1+\chi}
\end{equation}
with the Boltzmann weight $\chi\left(f\right) = \sqrt{k_\mathrm{g}/k_\mathrm{e}}\exp\left[-\beta\left(f^2/2\tilde{k} - f \Delta l + \epsilon\right)\right]$, the inverse temperature $\beta = 1/\left(k_\mathrm{B}T \right)$, and the auxiliary spring constant $\tilde{k} = k_\mathrm{g}k_\mathrm{e}/(k_\mathrm{e} - k_\mathrm{g})$.
Using this mean excitation, we can write the mean extension in the compact form
\begin{equation}
\label{eq:xNf}
\begin{split}
	\langle x_N\left(f\right) \rangle/N
	&= l_\mathrm{g} + p_\mathrm{eq}\left(f\right)\frac{f}{k_\mathrm{g}} + q_\mathrm{eq}\left(f\right)\left(\Delta l + \frac{f}{k_\mathrm{e}}\right).
\end{split}
\end{equation}

In this paper, we want to expand this model by introducing non-equilibrium kinetics.
The Hookean force $f_j\left(t \right) = k_{j+1} \left(x_{j+1} - x_{j} - l_{j+1}\right) - k_{j} \left(x_{j} - x_{j-1} - l_{j}\right)$ depends on the positions $x_j$ and the states of the springs, forming a coupled system of kinetics, elasticity, and state transitions. To solve our model numerically, we use Brownian dynamics (BD) computer simulations. In particular, for the inner springs $j=1,\cdots,N-1$ we use the overdamped Langevin equation \cite{Uhlenbeck_1930, AllenTildesley, DoiEdwards}.
\begin{equation}
	\label{eq:Langevin}
	\xi\dot{x}_j = k_{j+1} \left(x_{j+1} - x_{j} - l_{j+1}\right) - k_{j} \left(x_{j} - x_{j-1} - l_{j}\right) + \eta
\end{equation}
with friction constant $\xi$, Diffusion constant $D = 1/\xi \beta$ and white noise $\eta$ with zero mean, $\langle \eta \rangle = 0$, and which satisfies the fluctuation-dissipation theorem \cite{Callen_1951, Einstein_1905, Kubo_1966}
\begin{equation}
	\langle\eta\left(t\right) \eta\left(t'\right) \rangle = 2\xi k_\mathrm{B}T \delta\left(t-t'\right).
\end{equation}
The left end is fixed $x_0 = 0$ and the external force $f_{\rm ext}$ acts on the right edge, so that for the most right module, 
\begin{equation}
	\label{eq:LangevinRightEdge}
	\xi\dot{x}_N =  - k_{N} \left(x_{N} - x_{N-1} - l_{N}\right) + f_{\rm ext} + \eta.
\end{equation}

We integrate the coupled Langevin equations using a simple Euler-forward scheme \cite{AllenTildesley}. In addition to the positional dynamics, the internal states of the springs undergo stochastic transitions between the two energy branches. These transitions are implemented by a continuous-time Markov jump process -- specifically a Poisson two-state dynamics with Metropolis transition rates \cite{Metropolis_1953, Berthin_2025, Fries_2025} -- applied sequentially to each element at every simulation step.
For each spring, a potential state change is attempted with probability $P_\mathrm{trial} = 1 - e^{-\nu \Delta t}$, which defines the intrinsic attempt frequency and thus sets the kinetic time scale through the parameter $\nu$.

Once a trial is initiated, the acceptance of the new state follows the Metropolis criterion as a specific kinetic choice to ensure local detailed balance. Similar kinetically constrained approaches have recently been successfully employed to model the physical time evolution of driven soft matter systems \cite{Berthin_2025, Fries_2025}. The total transition probability therefore reads $P_\mathrm{acc} = P_\mathrm{trial} \min\!\left[1, e^{-\beta \Delta E}\right]$,
where $\Delta E$ is the energy difference between the proposed and the current state.
The first factor regulates how often transitions are attempted (kinetic prefactor), while the second factor enforces thermodynamic consistency by favoring energetically downward moves and suppressing upward ones according to the Boltzmann weight.

{All simulations are performed at a fixed thermal energy scale $k_\mathrm{B}T = \epsilon$ (i.e., $\beta = 1/\epsilon$). The integration time step is chosen as $\Delta t = 10^{-3} \, l_\mathrm{g}^2/D$, except for high driving frequencies $\omega \gtrsim 10 \, D/l_\mathrm{g}^2$, where it is reduced to $\Delta t = 10^{-4} \, l_\mathrm{g}^2/D$ to ensure numerical stability. Each run comprises an equilibration phase of $10^7$ steps followed by a production phase of $10^9$ steps. Within the stationary state, observables are averaged over multiple driving periods $T = 2\pi/\omega$ (i.e., over $t + zT$ with $z \in \mathbb{Z}$) to yield a single smooth period cycle from which extension, excitation $q$, and their phases are determined. We employ two distinct chain lengths depending on the phenomenon: $N=50$ springs are used to study global response properties like the frequency plateau (where the total chain length is comparable to the penetration depth), while $N=300$ springs are used to analyze spatial screening effects, ensuring the total chain length $N l_\mathrm{g}$ significantly exceeds the penetration depth $\lambda$ so that the force-influenced region remains small compared to the entire chain.}

\section{Continuum theory}

For very large $N$, we do not have to resolve each individual particle of the chain. Instead, we derive a continuum theory by coarse-graining the microscopic degrees of freedom. We introduce local fields for the excitation probabilities $q(X,t) \in [0,1]$ and $p(X,t) = 1-q(X,t)$, and define the displacement field $u(X) = x_j - j l_\mathrm{g}$ evaluated at the material coordinate $X = j l_\mathrm{g}$. In the transition from the discrete Langevin equation \eqref{eq:Langevin} to the continuous limit, the noise vanishes in the description because its contributions average out over the coarse-grained spatial and temporal scales.

Expressing the deterministic part of the discrete force balance directly in terms of the displacement $u_j$ and using finite differences, the equation of motion reads
\begin{equation}
\label{eq:discrete_u}
\xi\dot{u}_j = l_\mathrm{g} \frac{ k_{j+1} \left( l_\mathrm{g} \frac{u_{j+1} - u_j}{l_\mathrm{g}} - q_{j+1} \Delta l \right) - k_{j} \left( l_\mathrm{g} \frac{u_j - u_{j-1}}{l_\mathrm{g}} - q_{j} \Delta l \right) }{l_\mathrm{g}}.
\end{equation}
In the continuum limit ($j \to X$), the discrete differences converge to spatial derivatives, the spring constants become the continuous field $k_\mathrm{eff}(X)$, and the fraction becomes the spatial derivative of the enclosed term. Using $\xi = 1/(D\beta)$, we directly obtain the first continuum theory (CT) equation for the displacement:
\begin{equation}
\label{eq:ContinuousBistable}
\dot{u} = D\beta l_\mathrm{g} \, \partial_X \left( k_\mathrm{eff} \left( l_\mathrm{g} \partial_X u -  q \Delta l \right) \right).
\end{equation}
Here, the local effective stiffness $k_\mathrm{eff}(q)$ corresponds to the series arrangement of ground-state and excited springs with fractions $(1-q)$ and $q$, respectively:
\begin{equation}
\label{eq:keff}
k_\mathrm{eff}(q) = \frac{1}{\frac{1-q}{k_\mathrm{g}} + \frac{q}{k_\mathrm{e}}}.
\end{equation}
Equation \eqref{eq:ContinuousBistable} reveals that the chain motion is driven by the gradient of a local quantity. We can therefore identify the local elastic stress explicitly as
\begin{equation}
\label{eq:local_stress}
\sigma(X,t) = k_\mathrm{eff}(q) \left( l_\mathrm{g} \partial_X u - q \Delta l \right).
\end{equation}
This stress definition accounts for both the geometric strain $l_\mathrm{g} \partial_X u$ and the shift in the local equilibrium length due to the excited modules, given by $q \Delta l$. By introducing the field $q(X,t)$, we have introduced explicit spatial dependence into the parameters of the chain, as both $k_\mathrm{eff}$ and the unfolding term $q \Delta l$ are now functions of $X$.

Finding an update rule for $q(X,t)$ requires considering the microscopic mechanism of the state transitions, consistent with the Brownian dynamics simulations. While the simulations employ discrete state variables that switch stochastically according to a Metropolis algorithm, the continuous field $q(X,t)$ represents the local ensemble average of these binary states. Consequently, the time evolution of this mean field follows a master equation in the continuum limit, leading to the rate equation:
\begin{equation}
\label{eq:dq_new}
\frac{\mathrm{d}q}{\mathrm{d}t} = r_\mathrm{ge}\left(1-q \right) - r_\mathrm{eg}q
\end{equation}
with the effective forward and backward rates given by the harmonic sum of a diffusion-limited and a barrier-limited contribution:

\begin{equation}
\label{eq:rates_new}
\frac{1}{r_\mathrm{ge}} = \frac{1}{r_\mathrm{ge,B}} + \frac{1}{r_\mathrm{ge,\nu}},
\qquad
\frac{1}{r_\mathrm{eg}} = \frac{1}{r_\mathrm{eg,B}} + \frac{1}{r_\mathrm{eg,\nu}}.
\end{equation}
Here, $r_\mathrm{ge,\nu} \left(f\right) = \nu q_\mathrm{eq}\left(f\right)$ and $r_\mathrm{eg,\nu}\left(f\right) = \nu p_\mathrm{eq}\left(f\right)$ capture the diffusion-limited dynamics in the absence of a barrier, governed solely by the microscopic attempt frequency $\nu$ and the equilibrium probabilities $q_\mathrm{eq}\left(f\right)$ and $p_\mathrm{eq}\left(f\right) = 1 - q_\mathrm{eq}\left(f\right)$. The barrier-limited rates $r_\mathrm{ge,B}$ and $r_\mathrm{eg,B}$ are derived in appendix~\ref{sec:Berglund} using the framework of \cite{Berglund_2009, Chushnyakova_2020}. They read
\begin{equation}
	\label{eq:Berglund_rates}
\begin{split}
r_\mathrm{ge,B}
&= \nu \sqrt{\frac{k_\mathrm{g}}{2\pi\beta}} \left(\frac{\left|a_\mathrm{g}\right| + \left|a_\mathrm{e}\right|}{\left|a_\mathrm{g}a_\mathrm{e}\right|}\right) e^{-\beta\left(E^*-E_\mathrm{g} \right)}\\
r_\mathrm{eg,B}
&= \nu \sqrt{\frac{k_\mathrm{e}}{2\pi\beta}} \left(\frac{\left|a_\mathrm{g}\right| + \left|a_\mathrm{e}\right|}{\left|a_\mathrm{g}a_\mathrm{e}\right|}\right) e^{-\beta\left(E^*-E_\mathrm{e} \right)}
\end{split}
\end{equation}
{and consist of an exponential Arrhenius term \cite{Arrhenius_1889}, which depends on the barrier energy $E^*$ and the minima $E_\mathrm{g}$ and $E_\mathrm{e}$,  and a prefactor that accounts for the curvatures at the minima (spring constants $k_i$) and the slopes $a_i$ at the non-differentiable barrier (see Fig.~\ref{fig:energy_landscape}).}
This addition of inverse rates corresponds to a sequential process: the total transition time is the sum of the time required to reach the transition state via diffusion and the time to cross the barrier. In the limit of vanishing conformational change ($\Delta l \to 0$), the barrier disappears and the barrier limited rates diverge ($r_{\mathrm{\alpha\beta},\mathrm{B}} \to \infty$), leaving the dynamics purely diffusion-limited ($r_\mathrm{\alpha\beta} \approx r_\mathrm{\alpha\beta,\nu}$). Conversely, for large barriers, the crossing time dominates and we recover the Kramers-like limit ($r_\mathrm{\alpha\beta} \approx r_\mathrm{\alpha\beta,B}$).

These rates capture the energy landscape of the system, which is tilted by the local force $f = k_\mathrm{eff}l_\mathrm{g} \partial_X u$. The energy gaps between the ground and excited states are taken into account via
\begin{equation}
E_\mathrm{g} = -\left(\frac{f^2}{2k_\mathrm{g}} + f l_\mathrm{g}\right),
\qquad
E_\mathrm{e} = -\left(\frac{f^2}{2k_\mathrm{e}} + f l_\mathrm{e}\right) + \epsilon,
\end{equation}
which correspond to the minima located at
\begin{equation}
x_\mathrm{g} = l_\mathrm{g} + \frac{f}{k_\mathrm{g}},
\qquad
x_\mathrm{e} = l_\mathrm{e} + \frac{f}{k_\mathrm{e}}.
\end{equation}
The barrier energy $E^*=V_\mathrm{g}\left(x^*\right)=V_\mathrm{e}\left(x^*\right)$ is determined by the intersection point of the two {individual potentials $V_i = k_i/2\left(x-l_i\right)^2$} (see Fig.~\ref{fig:energy_landscape} for a schematic of the energy landscape and the relevant energy scales). Furthermore, the prefactor in the rates arises from the specific geometry of the potential well, characterized by the slopes
\begin{equation}
a_\mathrm{g} = k_\mathrm{g}(x^* - l_\mathrm{g}) - f,
\qquad
a_\mathrm{e} = k_\mathrm{e}(x^* - l_\mathrm{e}) - f
\end{equation}
at the barrier position $x^*$.
Since both the diffusion-limited and the barrier-limited rates independently satisfy detailed balance with the same ratio $\chi$, their combined effective rates also satisfy
\begin{equation}
\frac{r_\mathrm{ge}\left(f\right)}{r_\mathrm{eg}\left(f\right)} = \frac{q_\mathrm{eq}\left(f\right)}{p_\mathrm{eq}\left(f\right)} = \chi\left(f\right).
\end{equation}

Eqs. (\ref{eq:ContinuousBistable}) and (\ref{eq:dq_new}) with the just described definitions of kinetic and energy landscape parameters constitute our continuum theory, abbreviated CT in the following. Since we do not find analytic solutions for the CT, we rely on numerical results. Therefore, we update all $u\left(X,t \right)$ simultaneously by an implicit Euler algorithm. The equations of motion can be written in matrix form, where the occurring tridiagonal matrix guarantees efficient calculation of the new positions. For $q(X,t)$, a simple forward Euler is sufficient.

\subsection{Monostable limit and analytical solutions}

While the coupled dynamics of the general bistable case presented in the CT are only numerically accessible, the monostable limit admits analytical solutions, providing valuable physical insight and a baseline reference. The effective stiffness reduces to the constant ground-state value $k_\mathrm{eff}=k_\mathrm{g}$, and the change in equilibrium length $\Delta l$ vanishes. Consequently, the CT eq.~\eqref{eq:ContinuousBistable} simplifies to the diffusion equation
\begin{equation}
\label{eq:ContinuousMonostable}
\dot{u} = \frac{ k_\mathrm{g} l_\mathrm{g}^2}{\xi} \partial_X^2 u = \alpha \partial_X^2 u,
\end{equation}
where we have introduced the effective diffusion coefficient $\alpha = k_\mathrm{g} l_\mathrm{g}^2 / \xi$ to keep the expressions compact.
Mathematically, Eq.~\eqref{eq:ContinuousMonostable} is isomorphic to the classical 1D heat conduction equation for the displacement field $u(X,t)$, describing the diffusive propagation of mechanical deformation along the overdamped chain.

For a fixed left end $u(X=0, t) = 0$ and a periodic external force
\begin{equation}
\label{eq:BoundaryConditionPeriodicForce}
\left.k_\mathrm{g} l_\mathrm{g} \partial_X u\left(X,t \right)\right|_{X=L} = F_0\sin\left(\omega t \right),
\end{equation}
this linear partial differential equation can be solved analytically (cf. Appendix \ref{sec:Monostable}). Analytical solutions for such heat conduction initial-boundary value problems under periodic boundary conditions were derived in \citet{Xu_2023}; adapted to our mechanical continuum variables, the resulting monostable solution (Mono-CT) reads
\begin{widetext}
\begin{equation}
\begin{split}
\label{eq:PeriodicForceFinalSolution}
u\left(X,t \right)
&= \frac{F_0}{k_\mathrm{g}}X \sin\left(\omega t \right)
+ \frac{2F_0\omega\alpha}{Lk_\mathrm{g}}
\sum\limits_{n=1}^\infty \frac{\left(-1 \right)^{n}}{\left(\alpha \lambda_n^2 \right)^2 + \omega^2}
\left(\frac{\omega}{\alpha \lambda_n^2} \sin\left(\omega t \right) + \cos\left(\omega t \right) - e^{-\alpha \lambda_n^2 t}\right) \sin\left(\lambda_n X \right),
\end{split}
\end{equation}
\end{widetext}
with the discrete eigenvalues $\lambda_n=\frac{\left(2n-1 \right)\pi}{2L}$.

This solution can be interpreted as follows: The first term, $\frac{F_0}{k_\mathrm{g}}X \sin(\omega t)$, represents the direct, quasi-static mechanical response to the external drive. The sum over modes $n$ contains the dynamic corrections to this response. The sine and cosine terms within the summation account for the attenuation and the phase delay (time shift) of the propagating wave due to the finite relaxation time of the chain. Finally, the term proportional to $\exp(-\alpha \lambda_n^2 t)$ is a transient contribution, which decays exponentially and becomes irrelevant in the stationary state.

The modal structure of the solution reveals a strong dependence on the chain length $L = N l_\mathrm{g}$. The eigenvalues scale as $\lambda_n \sim 1/L \sim 1/N$. Consequently, the relaxation times of the individual modes, given by $\tau_n = 1/(\alpha \lambda_n^2)$, scale as $\tau_n \sim L^2 \sim N^2$. For the slowest mode ($n=1$), this implies that the characteristic time for the chain to reach its stationary state grows quadratically with the number of modules $N$, which is the hallmark of Rouse polymer dynamics~\cite{Rouse_1953, DoiEdwards}, where the slowest mechanical relaxation $\tau_1 = N^2 \xi/(\pi^2 k_\mathrm{g})$. 

\subsection{Timescales} 
The dynamics of the system involve two distinct relaxation processes with characteristic timescales: the mechanical relaxation of the spatial strain modes along the elastic backbone,

and the relaxation of the internal molecular states. With the spatial eigenvalues $\lambda_n$ derived in the previous section, each spatial mode relaxes with a characteristic rate $\alpha \lambda_n^2$. The slowest global mechanical relaxation mode corresponds to the fundamental spatial mode $n=1$:

\begin{equation}
    \label{eq:tau_mech}
    \tau_\mathrm{mech} \equiv \tau_1 = \frac{1}{\alpha \lambda_1^2} = \frac{4 N^2 \xi}{\pi^2 k_\mathrm{g}} = 4 \tau_\mathrm{Rouse}.
\end{equation}
Crucially, this fundamental relaxation time is four times larger than the classic Rouse time $\tau_\mathrm{Rouse}$ derived for free-ended polymer chains~\cite{Rouse_1953}. This factor of four originates directly from the fixed left boundary condition ($u(0,t) = 0$). While a free-ended chain permits a fundamental deformation mode spanning an antinode-to-antinode profile (a half-wavelength across the chain length $L$), clamping one end enforces a displacement node at $X=0$ and an antinode at the free end $X=L$. Consequently, the system length $L$ accommodates only a quarter-wavelength profile. This doubles the spatial wavelength of the fundamental mode and halves its spatial eigenvalue ($\lambda_1 = \frac{\pi}{2L}$). Because the relaxation rate scales quadratically with $\lambda_n$, this boundary constraint quadruples the slowest mechanical relaxation time.

The internal state dynamics follow a stochastic switching process, governed by an attempt frequency $\nu$. The acceptance probability depends on the instantaneous energy landscape, creating an effective, macroscopic state relaxation rate. In CT, we bridge the gap by explicit Kramers transition rates $r_\mathrm{ge}$ and $r_\mathrm{eg}$, as derived in Eq.~\eqref{eq:rates_new}. These rates contain the microscopic attempt frequency $\nu$ of the BD simulation, modulated by the geometric properties of the potential well. The exponential factors account for the probability of overcoming the energy barrier. Consequently, we obtain an effective timescale for the state relaxation that is not an independent phenomenological parameter, but emerges directly from the sum of the forward and backward rates:
\begin{equation}
	\label{eq:tau_q}
\tau_\mathrm{q} = \frac{1}{r_\mathrm{ge} + r_\mathrm{eg}}.
\end{equation}
This formulation captures the interplay between the two mechanisms: while the mechanical Rouse relaxation determines the background stress, the state relaxation time $\tau_\mathrm{q}$ is coupled to the mechanics through the force dependence of the rates. Thus, the CT accurately reproduces the physics of the BD simulation, where the transition probability depends implicitly on the local chain configuration.

\subsection{Linear Response Theory}
\label{sec:LinearResponse}

To characterize how the bistable chain responds to a small periodic external force $f_\mathrm{ext}$, we linearize the continuous theory around the stationary state at zero force. We consider deflections from equilibrium $\delta u(X,t) = u(X,t) - u_\mathrm{eq}(X)$ and $\delta q(X,t) = q(X,t) - q_\mathrm{eq}$, and neglect thermal noise. In the linear regime, the local change in stiffness due to excitation contributes only second-order terms, leaving $k_\mathrm{eff}^\mathrm{eq}$ as the only relevant stiffness.

Linearizing Eqs.~\eqref{eq:ContinuousBistable} and \eqref{eq:dq_new} and transforming to Fourier space ($\partial_t \rightarrow -i\omega$) yields the coupled equations
\begin{equation}
\begin{split}
-i\omega\tilde{u} &= D\beta k_\mathrm{eff}^\mathrm{eq} l_\mathrm{g} \, \partial_X \left( l_\mathrm{g} \partial_X \tilde{u} - \tilde{q} \Delta l\right),\\
-i\omega\tau_\mathrm{q} \tilde{q} &= -\tilde{q} + \frac{\gamma}{\Delta l} \left( l_\mathrm{g} \partial_X \tilde{u} - \tilde{q} \Delta l \right),
\end{split}
\label{eq:LinearizedCoupled}
\end{equation}
where we used the linearized susceptibility $\left. \mathrm{d}q / \mathrm{d}f \right|_{f=0} = \beta \Delta l \, q_\mathrm{eq}(1-q_\mathrm{eq})$ \cite{Pattloch_2025} and the bistability parameter $\gamma = \beta (\Delta l)^2 q_\mathrm{eq}(1-q_\mathrm{eq}) k_\mathrm{eff}^\mathrm{eq}$. {In the following, we refer to these equations and those derived from them as the linear response continuum theory (LRCT).}

The linearization significantly simplifies the coupling between mechanical and internal degrees of freedom: Eq.~\eqref{eq:LinearizedCoupled} allows us to solve algebraically for $\tilde{q}$ in terms of $\partial_X \tilde{u}$ and substitute it into the first equation. This decoupling yields a single differential equation for the displacement field alone:
\begin{equation}
\partial_X^2 \tilde{u} = - \frac{i\omega \xi}{k_\mathrm{eff}^\mathrm{eq} l_\mathrm{g}^2} \left(1+\frac{\gamma}{1-i\omega\tau_\mathrm{q}} \right) \tilde{u}.
\label{eq:DecoupledU}
\end{equation}
This is a diffusion equation with a complex diffusion coefficient. We introduce the complex wave vector $\tilde{k}$ (with dimensions of inverse length) via
\begin{equation}
\label{eq:ktilde_def}
\tilde{k}^2 = \frac{i\omega \xi}{k_\mathrm{eff}^\mathrm{eq} l_\mathrm{g}^2} \left(1+\frac{\gamma}{1-i\omega\tau_\mathrm{q}} \right).
\end{equation}
The detailed solution of this boundary value problem, including the application of boundary conditions at both chain ends, is provided in Appendix~\ref{app:LinearResponseDerivation}. The resulting linear response function reads
\begin{equation}
\label{eq:LinearResponse}
\chi_x(\omega) = \frac{\tilde{u}(L)}{\tilde{f}_\mathrm{ext}} = \frac{1}{k_\mathrm{eff}^\mathrm{eq} l_\mathrm{g} \tilde{k}} \left(1+\frac{\gamma}{1-i\omega\tau_\mathrm{q}}\right) \tan\left(\tilde{k} N l_\mathrm{g} \right).
\end{equation}
In the quasistatic limit $\omega \rightarrow 0$, we have $\tilde{k} \rightarrow 0$ and $\tan(\tilde{k} N l_\mathrm{g}) \approx \tilde{k} N l_\mathrm{g}$, yielding
\begin{equation}
\lim_{\omega\rightarrow 0} \frac{\tilde{\chi}_x(\omega)}{N} = \frac{1}{k_\mathrm{eff}^\mathrm{eq}} + \beta (\Delta l)^2 q_\mathrm{eq} (1-q_\mathrm{eq}),
\label{eq:QuasistaticLimit}
\end{equation}
which recovers the effective softness derived in \cite{Pattloch_2025}.

\subsection{Penetration Depth and Screening}
\label{sec:ScreeningLength}

The complex wave vector $\tilde{k}$ introduced in Eq.~\eqref{eq:ktilde_def} encodes both the spatial oscillation and the attenuation of mechanical signals. Its imaginary part determines the penetration depth $\lambda = 1/\mathfrak{Im}\tilde{k}$. For weak bistability ($\gamma \ll 1$), we expand the square root perturbatively to obtain the penetration depth (see Appendix~\ref{app:ScreeningDerivation} for details):
\begin{equation}
\label{eq:Penetration_depth}
\lambda \approx \sqrt{\frac{2 k_\mathrm{eff}^\mathrm{eq} l_\mathrm{g}^2}{\omega\xi}} \left(1 + \frac{1}{2} \frac{\gamma\left(1+\omega\tau_\mathrm{q} \right)}{1 + \omega^2\tau_\mathrm{q}^2}\right)^{-1}.
\end{equation}
This reveals the key physical effect: bistability enhances damping when the internal state relaxation is fast compared to the driving ($\omega \tau_\mathrm{q} \ll 1$). Here the correction factor approaches $1 + \gamma/2 > 1$, reducing $\lambda$ compared to the monostable case due to the additional dissipative channel from rapidly equilibrating state fluctuations. 

At high frequencies ($\omega \tau_\mathrm{q} \gg 1$), the correction vanishes and $\lambda$ recovers the monostable diffusive result, as the internal states cannot follow the rapid driving.

\section{Results}
\subsection{Dynamic response to periodic forcing}
\begin{figure}[h]
    \centering
        \includegraphics[width=0.45\textwidth]{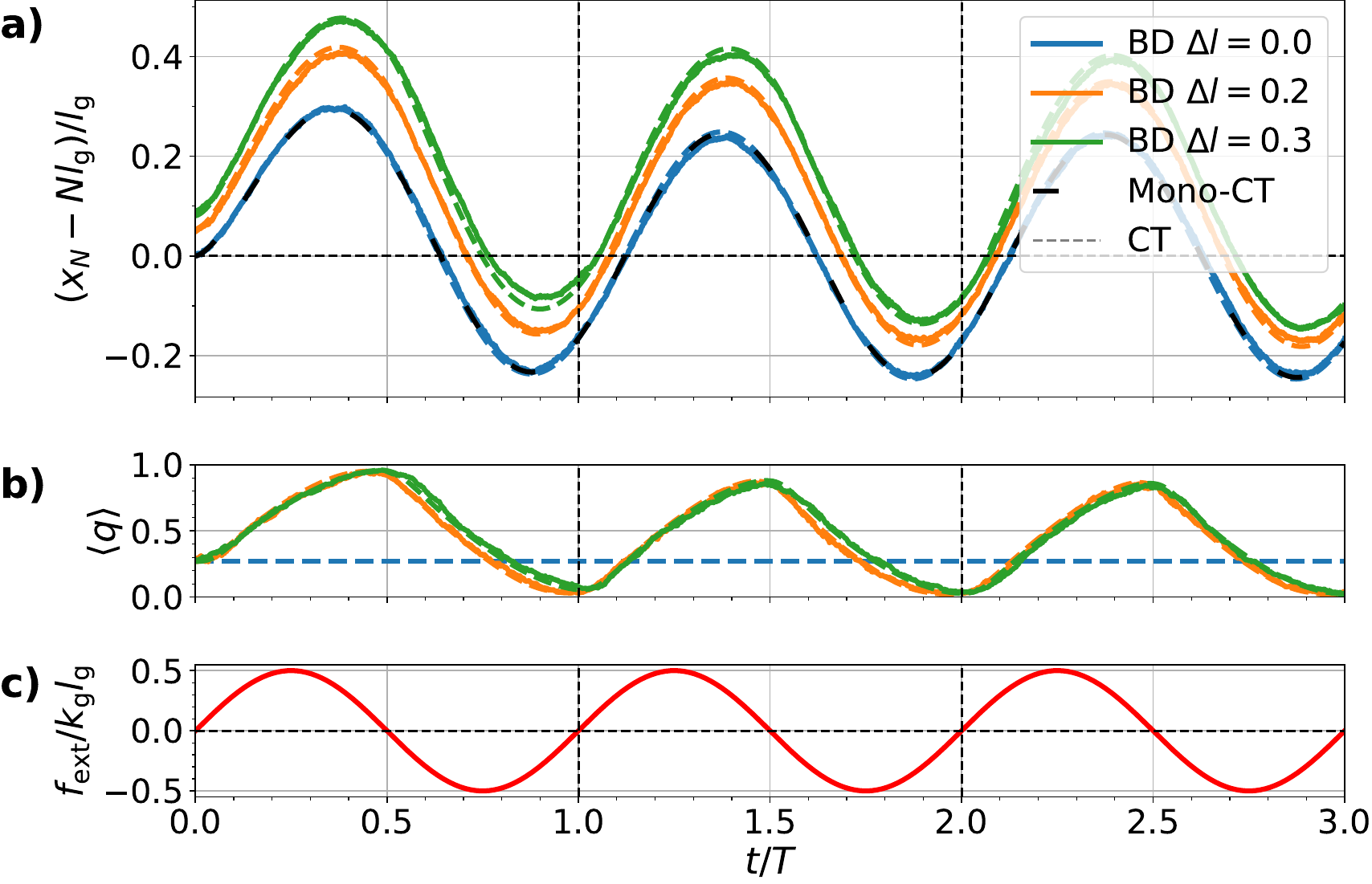}
    \caption{\textbf{Dynamic response of the bistable chain to periodic forcing.} Three panels show (\textbf{a}) normalized chain extension $(x_N-Nl_\mathrm{g})/l_\mathrm{g}$, (\textbf{b}) spatially averaged excitation fraction $\langle q \rangle(t) = \frac{1}{N}\sum_{i=1}^N q_i(t)$, and (\textbf{c}) applied force $f_\mathrm{ext}$. 
Solid lines represent Brownian dynamics (BD) simulations ($N=50$, $k_\mathrm{g}=100\,\epsilon/l_\mathrm{g}^2$, $\Delta k=0$, $\nu = 1\,D/l_\mathrm{g}^2$, $F_0=0.5\,k_\mathrm{g}l_\mathrm{g}$, $\omega=0.2\,D/l_\mathrm{g}^2$). Data are averaged over $M=10$ independent realizations with identical system parameters but different thermal noise to reduce stochastic fluctuations. 
Different colors correspond to $\Delta l/l_\mathrm{g}\in\{0,0.2,0.3\}$. 
The solid black line is the analytic monostable solution (Mono-CT) eq.~\eqref{eq:PeriodicForceFinalSolution}.
Dashed colored lines show numerical solutions of the CT, eq.~\eqref{eq:ContinuousBistable} and eq.~\eqref{eq:dq_new}.
Vertical black dashed lines mark integer multiples of the driving period $t/T = 1, 2, 3$. Horizontal black dashed lines indicate zero-reference levels: $(x_N-Nl_\mathrm{g})/l_\mathrm{g} = 0$, $\langle q \rangle = q_\mathrm{eq}$, and $f_\mathrm{ext} = 0$, respectively. Major grid lines are included to highlight the subtle differences in amplitude between the initial transient cycles and the symmetric stationary state.
\label{fig:periodic_response}
}
\end{figure}

Fig.~\ref{fig:periodic_response} demonstrates the dynamic response of the bistable spring chain to periodic forcing at the right boundary. 
After equilibration at zero force, a sinusoidal drive $f_\mathrm{ext}(t)=F_0\sin(\omega t)$ with amplitude $F_0=0.5\,k_\mathrm{g}l_\mathrm{g}$ and frequency $\omega=0.2\,D/l_\mathrm{g}^2$ is applied at $t=0$. 
The system is simulated over three periods ($T=2\pi/\omega$) using BD simulations with $N=50$ springs, where results are averaged over $M=10$ independent realizations to obtain smooth curves.

Both the total chain extension $L(t)$ and the mean excitation fraction $q(t)$ faithfully follow the driving force but with a characteristic phase lag. 
Fig.~\ref{fig:periodic_response}\textbf{a}) shows the normalized extension $(L-Nl_\mathrm{g})/l_\mathrm{g}$, which oscillates around its mean value with the same frequency $\omega$ as the external force (panel \textbf{c})). Interestingly, the response during the initial cycles is slightly asymmetric. Because no force is applied prior to $t=0$, the system undergoes a switch-on transient, which is explicitly captured by the exponential decay terms $\mathrm{e}^{-\alpha\lambda_n^2t}$ in the analytic monostable solution, eq.~\eqref{eq:PeriodicForceFinalSolution}.

Physically, this initial asymmetry stems from a transient switch-on effect: during the first period, the slow modes are largely in their equilibrium state due to the lack of prior compression, allowing the fast modes to dominate the response. In subsequent periods, the slow modes remain slightly compressed, leading to a reduced amplitude. By the third period, these transients have largely decayed, and the oscillation reaches its symmetric stationary state. For the bistable cases ($\Delta l>0$), panel \textbf{b}) reveals that $q(t)$ similarly traces a sinusoidal trajectory, phase-shifted relative to both the force and the mechanical response.

Remarkably, all three approaches -- BD simulations (solid), CT (dashed), and the Mono-CT ($\Delta l=0$) eq.~\eqref{eq:PeriodicForceFinalSolution} -- show very good agreement across the full period. 
This validates both (i) CT as an efficient surrogate for the microscopic BD dynamics, and (ii) the accuracy of the Mono-CT \eqref{eq:PeriodicForceFinalSolution} as a benchmark. 
The phase lags observed in $L(t)$ and $q(t)$ reflect the combined overdamped mechanical relaxation of the chain and the kinetic bistability of individual springs, with coupling between mechanical deformation and internal state transitions generating the observed non-trivial dynamics.

\subsection{Linear response: Bode plots}
\begin{figure*}[t]
\noindent\includegraphics[width=\textwidth]{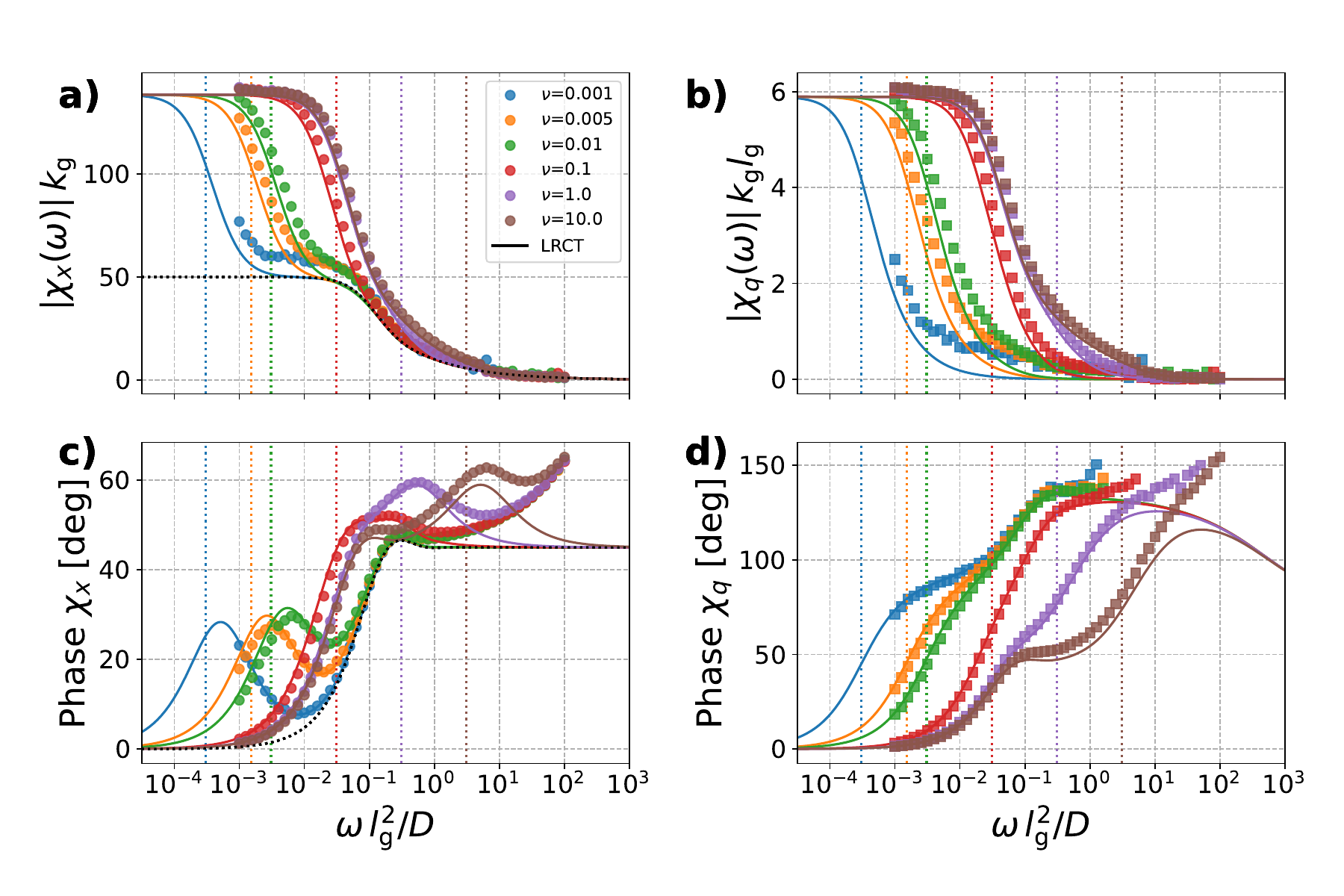}
\caption{
\textbf{Bode plots of the linear response.}
Panels \textbf{a}), \textbf{b}) show amplitude and \textbf{c}), \textbf{d}) phase of the chain extension $\chi_x(\omega)$ and mean excitation fraction $\chi_q(\omega)$, respectively.
Parameters are identical to Fig.~\ref{fig:periodic_response} except $F_0=0.1\,k_\mathrm{g}l_\mathrm{g}$ (small-amplitude regime) and $\Delta l/l_\mathrm{g}=0.3$.
Symbols represent BD simulations for which different colors denote distinct Metropolis rates $\nu$.
Solid lines show predictions from LRCT,  cf. eq.~\eqref{eq:LinearResponse}.
The black dashed line represents the monostable reference case without bistability ($\Delta l = 0$).
For the presented parameters, the mechanical Rouse relaxation time is $\tau_\mathrm{mech} = 2.53\,l_\mathrm{g}^2/D$, while the state relaxation time evaluates to $\tau_\mathrm{q} = 3.25/\nu$.
Vertical dotted lines indicate the characteristic frequencies $1/\tau_\mathrm{mech}$ (black) and $1/\tau_\mathrm{q}$ (color-coded to match the respective curves).
\label{fig:bode_plots}
}
\end{figure*}

\begin{figure*}[t]
\noindent\includegraphics[width=\textwidth]{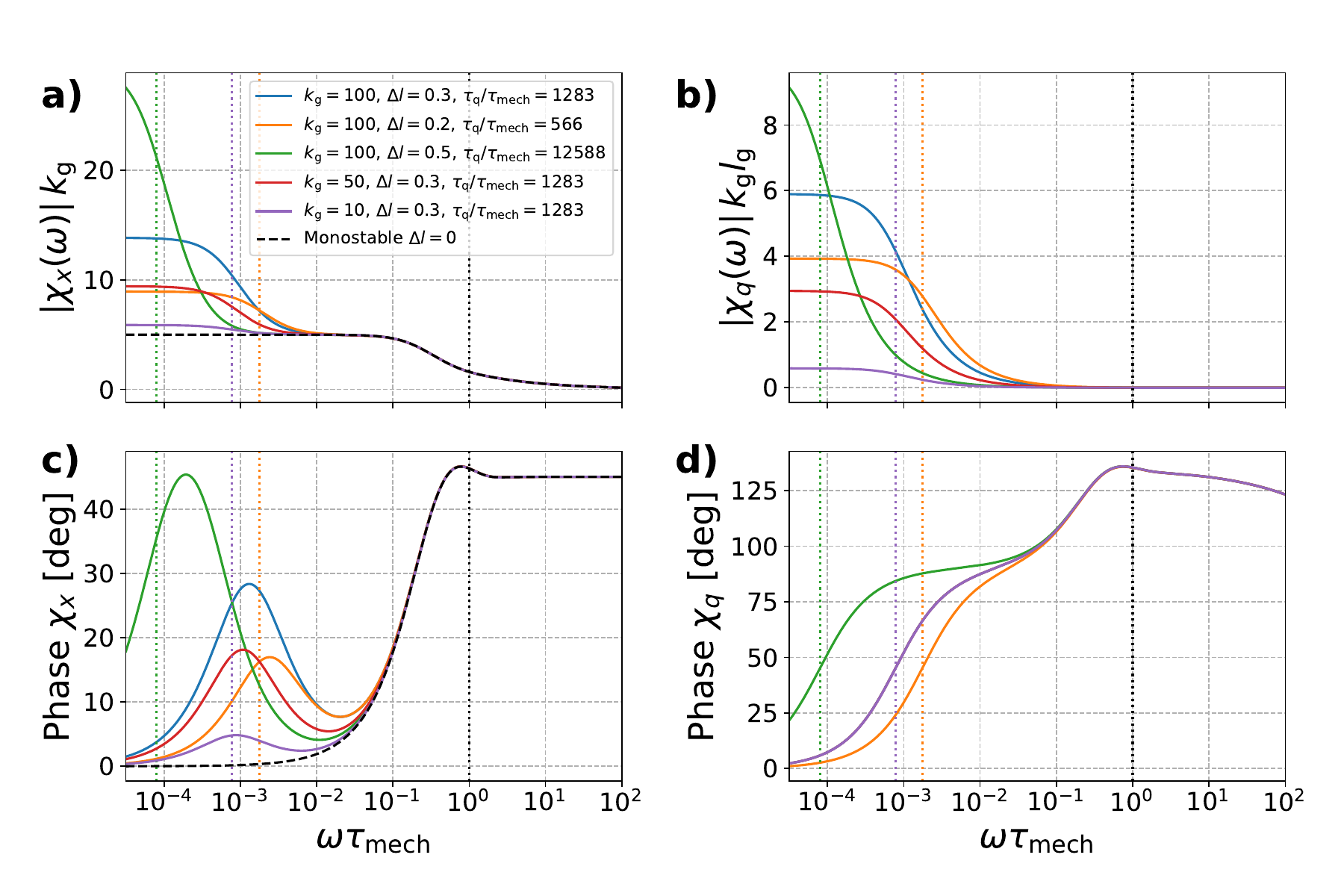}
\caption{
\textbf{Bode plots scaled with $\tau_\mathrm{mech}$.}
Panels \textbf{a}), \textbf{b}) show amplitude and \textbf{c}), \textbf{d}) phase of the chain extension $\chi_x(\omega)$ and mean excitation fraction $\chi_q(\omega)$, respectively.
All curves are calculated using LRCT (Eq.~\eqref{eq:LinearResponse}), which provides an accurate analytical description in this small-force regime ($F_0 = 0.1\,k_\mathrm{g}l_\mathrm{g}$).
The blue curve corresponds to the reference parameters from Fig.~\ref{fig:bode_plots} ($\nu = 1.0\,D/l_\mathrm{g}^2$, $\Delta l/l_\mathrm{g} = 0.3$).
For the orange and green curves, we vary only the conformational length $\Delta l$, which modifies the state relaxation time $\tau_\mathrm{q}$ while keeping the attempt frequency constant at $\nu = 1\,D/l_\mathrm{g}^2$.
For the red and purple curves, we vary the spring constant $k_\mathrm{g}$ and rescale the frequency axis by the mechanical relaxation time $\tau_\mathrm{mech}$(Eq.~\eqref{eq:tau_mech}, causing these curves to collapse onto each other. Simultaneously, we adjust $\nu$ to maintain a constant state relaxation time $\tau_\mathrm{q}$.
The black dashed line represents the monostable reference case ($\Delta l = 0$).
Vertical dotted lines indicate the characteristic frequencies $1/\tau_\mathrm{mech}$ (black) and $1/\tau_\mathrm{q}$ (color-coded to match the respective curves).
The plateau in $\chi_x$ emerges in the frequency gap between these two timescales, where mechanical relaxation is complete but state transitions have not yet activated.
\label{fig:bode_rescaled}
}
\end{figure*}

\begin{figure}[h]
\centering
\includegraphics[width=\columnwidth]{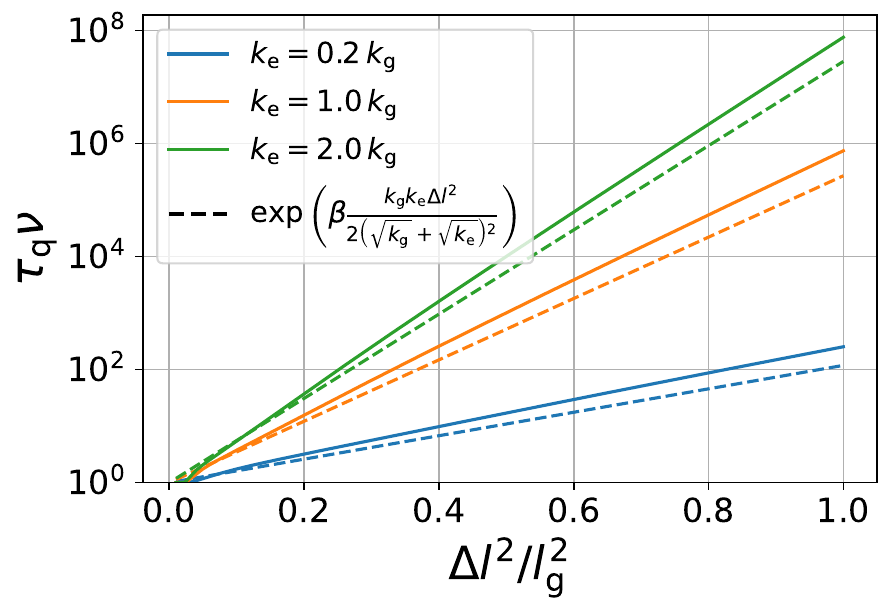}
\caption{
\textbf{State relaxation time $\tau_\mathrm{q}$ as a function of the conformational length $\Delta l$.} 
The relaxation time is plotted against $\Delta l^2$ to reveal the exponential scaling behavior. 
Solid lines show the exact values of $\tau_\mathrm{q}$ computed from the full transition rates according to Eq.~\eqref{eq:tau_q} for fixed ground-state stiffness $k_\mathrm{g} = 100\,\epsilon/l_\mathrm{g}^2$ and varying excited-state stiffness $k_\mathrm{e}$. 
Different colors correspond to different stiffness ratios: blue ($k_\mathrm{e} = 0.2\,k_\mathrm{g}$), orange ($k_\mathrm{e} = k_\mathrm{g}$), and green ($k_\mathrm{e} = 2.0\,k_\mathrm{g}$). 
Dashed lines represent the approximation for large spring constants, Eq.~\eqref{eq:tau_approx}, which captures the dominant exponential dependence on the barrier height. 
The plot demonstrates that $\tau_\mathrm{q}$ grows exponentially with $\Delta l^2$, as the conformational change directly determines the energy barrier that must be overcome during state transitions. 
\label{fig:tau_q_vs_Dl_squared}
}
\end{figure}

Fig.~\ref{fig:bode_plots} validates the LRCT by comparing BD simulations with analytical predictions across a wide range of driving frequencies $\omega$ at stationary conditions. To ensure small-amplitude conditions required by eq.~\eqref{eq:LinearResponse}, the force amplitude is reduced to $F_0=0.1\,k_\mathrm{g}l_\mathrm{g}$.  Amplitudes $\chi_x=A_x/F_0$ and $\chi_q=A_q/F_0$ (with $A_x=\max|L-L_\mathrm{eq}|$, $A_q=\max|q-q_\mathrm{eq}|$) are extracted from period-averaged signals over many cycles $t\in[0,T]$.

The mechanical response amplitude $\chi_x(\omega)$ exhibits three frequency regimes governed by the ratio $\tau_\mathrm{q}/\tau_\mathrm{mech}$. At high frequencies ($\omega \tau_\mathrm{mech} \gg 1$), the response vanishes as the chain cannot follow the drive. At low frequencies ($\omega \tau_\mathrm{q} \ll 1$), both mechanical and internal degrees of freedom equilibrate. Here, bistability enhances the response compared to the monostable case: state transitions generate additional length through unfolding, effectively stretching the response curve in the amplitude direction. For sufficiently separated timescales ($\tau_\mathrm{q} \gg \tau_\mathrm{mech}$), an intermediate plateau emerges where mechanical relaxation is complete ($\omega \tau_\mathrm{mech} \ll 1$) but internal states remain frozen ($\omega \tau_\mathrm{q} \gg 1$). In this regime, $\chi_x$ matches the monostable response since state transitions are kinetically suppressed.

Importantly, this plateau is a finite-size effect requiring $Nl_\mathrm{g} \lesssim \lambda$ such that mechanical relaxation spans the entire chain while internal transitions remain frozen. Fig.~\ref{fig:bode_rescaled} illustrates this scaling behavior using LRCT (which accurately reflects the microscopic BD dynamics in this small-force regime) by rescaling the frequency axis with $\tau_\mathrm{mech}$: the plateau occupies the gap between the characteristic frequencies $1/\tau_\mathrm{mech}$ and $1/\tau_\mathrm{q}$ (vertical dotted lines). The plateau width is thus controlled by the ratio $\tau_\mathrm{q}/\tau_\mathrm{mech}$.

The upper boundary of the plateau is dictated solely by mechanical properties. As shown in Fig.~\ref{fig:bode_rescaled}, rescaling the frequency by $\tau_\mathrm{mech} \propto k_\mathrm{g}^{-1}$ causes right ends of the curves to collapse perfectly, confirming that the high-frequency limit scales with $\omega \propto k_\mathrm{g}$. Crucially, for curves with the same $\tau_\mathrm{q}$ (red and purple), also the left end of the plateau collapses, demonstrating that the lower boundary is set by $\tau_\mathrm{q}$ independent of the specific mechanical parameters. However, $\tau_\mathrm{q}$ itself depends sensitively on both the attempt frequency $\nu$ and the barrier height $E^*$. This strong parameter dependence is illustrated in Fig.~\ref{fig:tau_q_vs_Dl_squared}, which shows $\tau_\mathrm{q}$ as a function of $\Delta l^2$ for various stiffness contrasts. For large spring constants and moderate $\Delta l$, the barrier position simplifies (for $k_\mathrm{g} = k_\mathrm{e}$ it is $x^* \approx l_\mathrm{g} + \Delta l/2$), yielding the approximation

\begin{equation}
\label{eq:tau_approx}
	\tau_\mathrm{q} \approx \frac{1}{\nu} \exp\left(\beta \frac{k_\mathrm{g}k_\mathrm{e}\Delta l^2}{2 \left(\sqrt{k_\mathrm{g}} + \sqrt{k_\mathrm{e}} \right)^2} \right).
\end{equation}

Increasing the barrier via larger $\Delta l$ thus slows down transitions dramatically, widening the plateau by pushing the onset of state switching to lower frequencies. This exponential dependence underlies the \emph{state-locking} phenomenon discussed in Section~\ref{sec:Parameter_study}: for large $\Delta l$, the barrier becomes so high that $\tau_\mathrm{q}$ exceeds the simulation timescale and transitions are effectively suppressed, a behavior reminiscent of energy-dominated locking mechanisms in rubber-like gels~\cite{Duarte_2023}. In the unfolding regime ($\omega < \tau_\mathrm{q}^{-1}$), the curves do not collapse because the bistable enhancement differs: larger $\Delta l$ yields more pronounced effects, and higher spring constants amplify the response since small forces induce larger changes in the equilibrium distribution. At very low frequencies, the static response amplitude increases with $k_\mathrm{g}$ through this unfolding mechanism.

The mechanical phase $\phi_x(\omega)$ exhibits non-monotonic behavior with peaks at $\omega \tau_\mathrm{q} \approx 1$ (internal transitions, colored vertical dotted lines in Fig.~\ref{fig:bode_plots}) and $\omega \tau_\mathrm{mech} \approx 1$ (mechanical relaxation, black vertical dotted line), producing a double-peaked structure for well-separated timescales (e.g., orange curve in Fig.~\ref{fig:bode_plots}). CT and BD simulations agree excellently up to intermediate frequencies. At extremely high frequencies ($\omega \gg 1/\tau_\mathrm{mech}$), LRCT deviates from BD ($90^\circ$ limit) because the continuum approximation $\lambda \gg l_g$ breaks down: the discrete system responds only at the boundary while CT enforces exponential decay throughout the chain.

The excitation amplitude $\chi_q(\omega)$ follows a strictly monotonic trend from $q_\mathrm{eq}(F_0)$ at $\omega \to 0$ to zero at high frequencies. Reducing $\nu$ shifts the curves to lower frequencies, freezing the degrees of freedom earlier. However, the inflection point does not scale linearly with $\nu$, revealing mechanical coupling: even for $\nu \gg 1$, the excitation response is limited by the chain's ability to transmit force to trigger switches.

The excitation phase $\phi_q(\omega)$ consistently lags behind $\phi_x$, revealing mechanical diffusion as an additional limiting factor. For rapid switching ($\nu=10\,D/l_\mathrm{g}^2$, brown), $\phi_q$ forms a plateau around $\omega \sim 0.1\,D/l_\mathrm{g}^2$ where state relaxation, not kinetics, limits chain extension. Only at $\omega \gtrsim 10\,D/l_\mathrm{g}^2$ does $\nu$ become the bottleneck, allowing $\phi_q$ to approach $\pi$. For slow rates ($\nu=0.001\,D/l_\mathrm{g}^2$, blue), timescale separation produces a double-peaked structure corresponding to distinct mechanical and internal relaxation timescales.

BD simulations and LRCT show excellent quantitative agreement for both amplitudes and phases across different $\nu$, provided $\lambda \gg l_\mathrm{g}$. Minor shifts arise from residual nonlinearities or thermal noise in BD. The force amplitude $F_0=0.1\,k_\mathrm{g}l_\mathrm{g}$ balances low-$\omega$ accuracy against high-$\omega$ noise while maintaining linear response conditions.

\subsection{Spatial penetration depth}
\begin{figure}[h]
\centering
\includegraphics[width=\columnwidth]{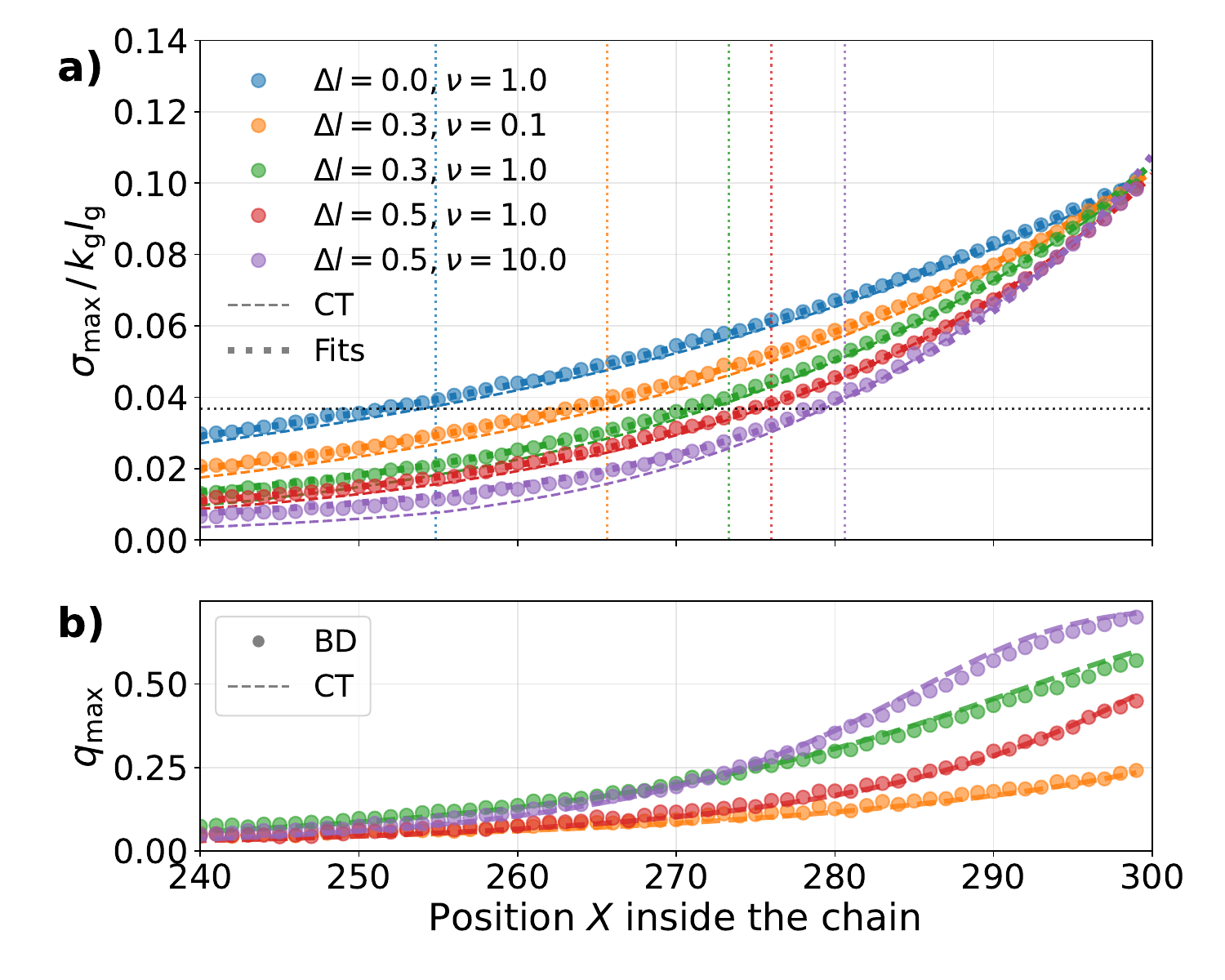}
\caption{\textbf{Spatial profiles of stress and excitation and the determination of the penetration depth $\lambda$.} \textbf{a}) shows the spatial profile of the maximum stress $\sigma_{\rm max}(X) = \max_t \sigma(X,t)$, decaying exponentially from the driven right end towards the fixed left end. The penetration depth $\lambda$ is extracted from an exponential fit $A\exp(-(L-X)/\lambda)+c$ (dotted line, shown only in this panel) and defined as the distance from the right boundary where the stress drops to $1/e$ of the driving amplitude $F_0$ (indicated by horizontal dotted lines). The fit agrees so well with the BD simulation data that the dotted line lies mostly within the BD symbols and is barely visible. \textbf{b}) displays the maximum deviation of the excitation fraction $q_{\rm max}(X)$ from equilibrium. Symbols: BD simulations (circles); dashed line: CT. Note that this analysis is conducted in the small-force regime ($F_0 = 0.1\,k_\mathrm{g}l_\mathrm{g}$), where LRCT yields virtually identical decay profiles; LRCT curves are omitted here solely to avoid visual clutter. Parameters are identical to Fig.~\ref{fig:bode_plots} with $\omega = 0.1\,l_\mathrm{g}^2/D$, except for the increased chain length $N=300$, which suppresses finite-size effects.}
\label{fig:lambda_determination}
\end{figure}

\begin{figure}[h]
\centering
\includegraphics[width=\columnwidth]{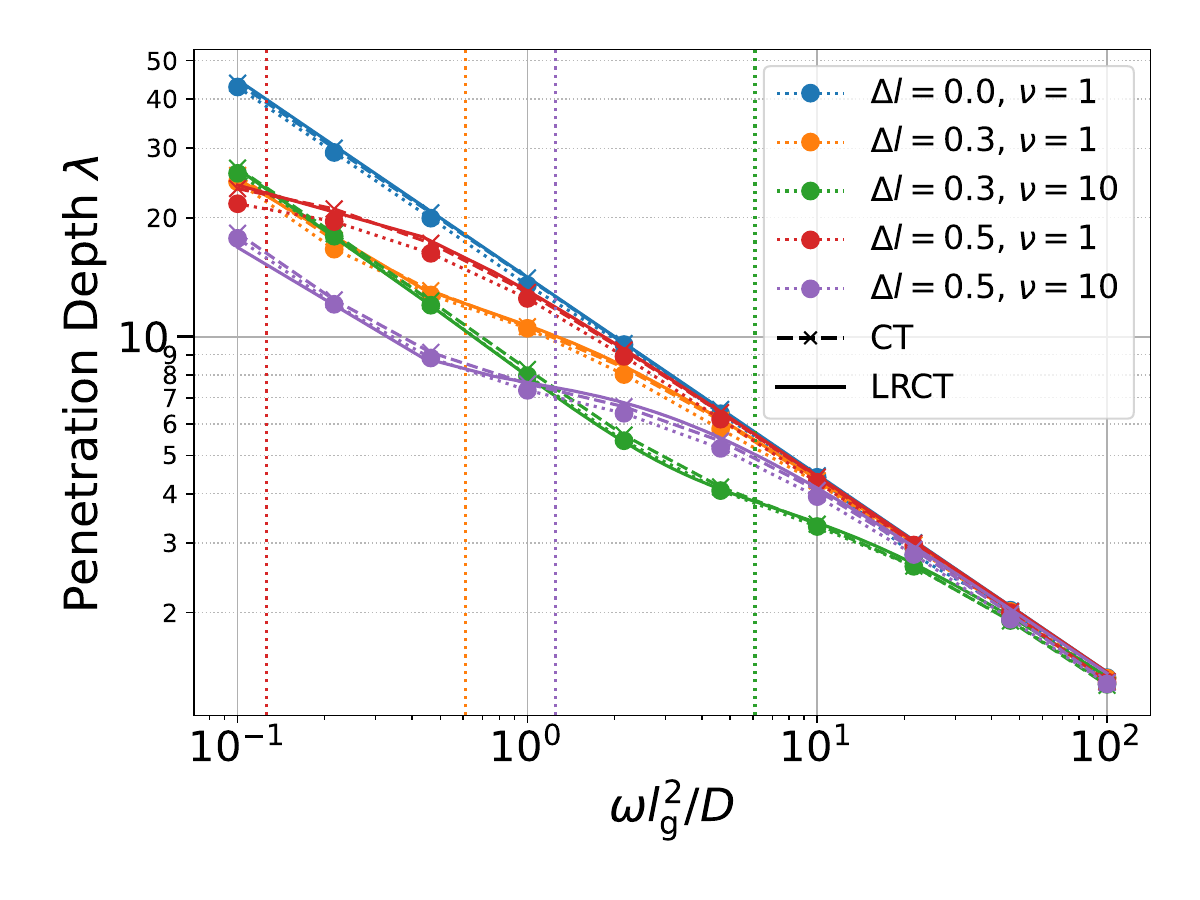}
\caption{
\textbf{Frequency-dependent penetration depth $\lambda(\omega)$.} 
Log-log plot for $N=300$ springs (same parameters as Fig.~\ref{fig:bode_plots}, $F_0=0.1\,k_\mathrm{g}l_\mathrm{g}$). 
Circles with dotted lines: BD simulations; dashed lines: CT; 
solid lines: analytical theory eq.~\eqref{eq:Penetration_depth}. 
\label{fig:penetration_depths}
}
\end{figure}

Having validated the LRCT in frequency space, we now examine its \emph{spatial} structure. While the results in the preceding section characterized the global response of the entire chain, we now consider the spatially-resolved analysis of local stress $\sigma(X)$ and excitation $q(X)$. 

Due to overdamping, we expect the signal to suffer exponential decay from the driven right boundary ($X=L$) towards the fixed left end. In the framework of LRCT, we calculated a complex wavevector $\tilde{k}$, which is independent of the spatial coordinate for small forces. This constant wavevector governs the spatial dependence of the solution, naturally leading to a strict exponential decay $\sigma(X) \sim e^{-X/\lambda}$. In contrast, in the nonlinear CT, the nonlinear coupling induces position-dependent wavevectors, resulting in a decay profile that can deviate from a strict exponential form.
Fig.~\ref{fig:lambda_determination} illustrates how the penetration depth $\lambda$ is rigorously extracted from the spatial profiles in both simulation approaches. Panel \textbf{a}) depicts $\sigma_{\rm max}(X) = \max_t \sigma(X,t)$, showing the expected exponential decay. We determine $\lambda$ by fitting an exponential envelope to this profile (dashed line), defining $\lambda$ as the distance from the right boundary where the stress amplitude decays to $1/e$ of the driving amplitude $F_0$. The corresponding profile for the excitation fraction $q_{\rm max}(X) = \max_t (q(X,t) - q_{\mathrm{eq}})$ is shown in panel \textbf{b}).
With the exact transition rates employed in the continuous model, the $q_{\rm max}$ profiles obtained from BD simulations and from the CT overlap virtually perfectly for all values of $\nu$. While the stress profiles $\sigma_{\rm max}(X)$ from BD exhibit a slight offset (marginally elevated stress levels by noise), this does not affect the extracted penetration depth: the fitted decay lengths $\lambda$ from BD and CT agree very well. Consequently, $\lambda$ provides a consistent metric to quantify signal penetration for both BD and CT data.

The analysis presented in the following focuses on the small-force regime ($F_0 = 0.1\,k_\mathrm{g}l_\mathrm{g}$) to allow for a direct comparison with linear response theory. Furthermore, we restrict our study here to $\Delta k = 0$, as the dissipative effects are dominated by the length change $\Delta l$. Our linear response formulation (Eq.~\eqref{eq:Penetration_depth}) corroborates this, showing that $\Delta k$ contributes only via the effective stiffness $k_{\rm eff}^{\rm eq}$ without inducing qualitative changes in the scaling of $\lambda(\omega)$. Simulations with large forces (e.g. Fig.~\ref{fig:penetration_vs_Dl_Dk_large_force}) further confirm that the screening behavior remains essentially governed by $\Delta l$. Notably, for small forces, the system exhibits a symmetry with respect to the sign of $\Delta l$; the coupling parameter $\gamma$ enters quadratically ($\gamma \propto (\Delta l)^2$), implying that dissipation by state transitions is independent of the direction of deformation. This symmetry is broken only in the non-linear regime, which can lead to a second transition for $\Delta k \neq 0$—similar to the static case reported in \cite{Pattloch_2025}—and a non-monotonic shape of $q_{\rm max}(X)$. However, even this dynamic implementation of the second transition does not alter the qualitative scaling of the mechanical screening length $\lambda(\omega)$.

With $\lambda$ established as a robust metric, Fig.~\ref{fig:penetration_depths} reveals the dramatic effect of bistability on wave propagation. The monostable case (blue) follows the classical $\lambda \propto \omega^{-1/2}$ scaling throughout. Bistable cases (orange, green, purple, red) exhibit a characteristic \emph{kink}: at intermediate frequencies $\lambda$ drops below the monostable baseline before recovering the $\omega^{-1/2}$ asymptote at both frequency extremes.

Two further consequences of increasing the bistability parameter $\gamma$ (here by increasing $\Delta l$) become evident. First, a larger $\Delta l$ raises the mechanical contrast, thereby strengthening the screening (attenuation) effects relative to the monostable case. Second, the same increase raises the energy barrier that must be overcome during a state transition. Because crossing a higher barrier requires more time, the onset of the screening regime---the kink in $\lambda(\omega)$---shifts to lower frequencies for a fixed transition rate $\nu$ (compare red vs. orange or purple vs. green curves). In other words, the crossover from the bistable to the damped response occurs earlier in the frequency spectrum when the barrier is higher. If $\Delta l$ is pushed beyond a critical value, the barrier becomes effectively insurmountable on the simulation timescale, leading to \emph{state-locking}. In this limit the chain remains frozen in a single state and the response reverts to the monostable $\lambda \propto \omega^{-1/2}$ behavior.

To pinpoint the onset of bistability-enhanced screening, we define the inflection point via the logarithmic curvature
\begin{equation}
\label{eq:TurningPoint}
\frac{\mathrm{d}^2\log\lambda}{\mathrm{d}(\log\omega)^2} = 0.
\end{equation}

Direct numerical evaluation by eq.~\eqref{eq:Penetration_depth} is algebraically cumbersome, so we use the alternative small-$\gamma$ approximation

\begin{equation}
	\label{eq:exp_penetration}
	\lambda \approx \sqrt{\frac{2 k_\mathrm{eff}^\mathrm{eq}}{\omega\xi}} 
	\exp\left[-\frac{\gamma}{2}\frac{1+\omega\tau_\mathrm{q}}{1 + \omega^2\tau_\mathrm{q}^2}\right],
\end{equation}
which is more convenient because of the outer exponential function.
The relevant positive root of eq.~\eqref{eq:TurningPoint} occurs at $\omega\tau_\mathrm{q} =-1 + \sqrt{2} + \sqrt{4-2\sqrt{2}}\approx 1.50$, demonstrating that the screening onset scales universally with the ratio of driving frequency to internal relaxation time. Faster switching ($\nu \uparrow$, $\tau_\mathrm{q} \downarrow$) thus shifts screening to higher frequencies, as observed.

\emph{Connection between Response Plateau and Screening:} The frequency-insensitive response plateau and the enhanced spatial screening are complementary manifestations of the same underlying timescale separation. As detailed in the previous sections, both the plateau's emergence and the screening kink at $\omega \tau_{\mathrm{q}} \approx 1.5$ are governed by the relative positioning of $\tau_{\mathrm{q}}$ and $\tau_{\mathrm{mech}}$. This establishes a unified design principle: tuning the internal kinetics relative to mechanical relaxation simultaneously controls both the frequency range of stable compliance and the strength of spatial attenuation.

These findings reveal that the chain's frequency-dependent attenuation characteristics are defined by the coupling between its mechanical rigidity and its internal kinetics. To effectively deploy this mechanism for specific applications, one must coordinate these two timescales. In the subsequent section, we translate this understanding into explicit design guidelines for tailoring such materials to predefined attenuation and compliance requirements.

\subsection{Parameter study}
\label{sec:Parameter_study}

\begin{figure}[h]
\centering
\includegraphics[width=\columnwidth]{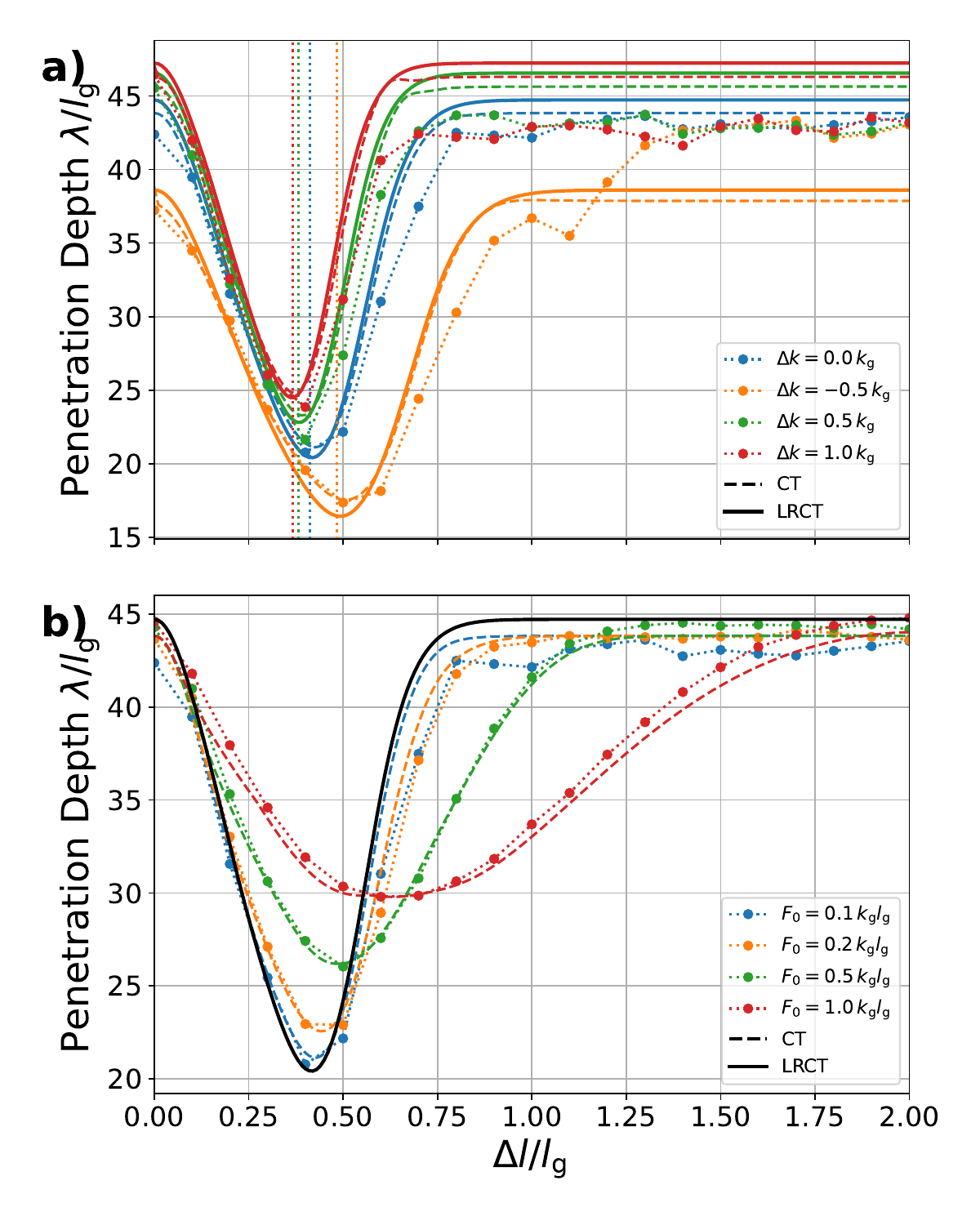}
\caption{
\textbf{Dependence of the penetration depth on the bistable length $\Delta l$ and varying parameters.}
The penetration depth $\lambda$ is plotted against $\Delta l$ for (\textbf{a}) varying stiffness contrast $\Delta k \in \{-0.5, 0.0, 0.5, 1.0\}\,k_\mathrm{g}$ at fixed force $F_0 = 0.1\,k_\mathrm{g}l_\mathrm{g}$, and (\textbf{b}) varying driving amplitude $F_0$ at fixed stiffness contrast $\Delta k = 0$.
Common fixed parameters are $\omega = 0.1\,D/l_\mathrm{g}^2$, $k_\mathrm{g} = 100\,\epsilon/l_\mathrm{g}^2$, and $\nu = 1.0\,D/l_\mathrm{g}^2$.
The blue curve represents the reference case ($\Delta k = 0$, $F_0 = 0.1\,k_\mathrm{g}l_\mathrm{g}$) and is identical in both panels.
Symbols denote BD simulations, dashed lines continuum theory (CT), and solid lines linear response theory (LRCT) according to eq.~\eqref{eq:Penetration_depth}.
Vertical dotted lines indicate the resonance condition $\omega \tau_\mathrm{q} = 1$.
\label{fig:penetration_vs_Dl_combined}
}
\end{figure}

To understand how the bistable material parameters control the wave propagation, we systematically vary the conformational length $\Delta l$, the stiffness contrast $\Delta k$ and the driving amplitude $F_0$ in Fig.~\ref{fig:penetration_vs_Dl_combined}. 

Fig.~\ref{fig:penetration_vs_Dl_combined}\textbf{a}) shows the penetration depth $\lambda$ as a function of $\Delta l$ for a small driving force $F_0 = 0.1\,k_\mathrm{g}l_\mathrm{g}$. Generally, we observe a non-monotonic behavior: for tiny $\Delta l$, bistability is barely noticeable and the system behaves like the monostable limit. As $\Delta l$ increases, the bistable screening effects become dominant and the penetration depth decreases significantly, reaching a distinct minimum. However, beyond this minimum, further increasing $\Delta l$ raises the energy barrier so much that state transitions are effectively suppressed. Due to this \emph{state locking}, the system reverts to an effectively monostable behavior and the additional screening vanishes, causing $\lambda$ to increase again.

The stiffness contrast $\Delta k$ strongly modifies this picture. A softer excited state (lower $k_\mathrm{e}$, i.e., more negative $\Delta k$) generally leads to a lower penetration depth, as the softer springs facilitate excitation and softer springs inherently transmit mechanical signals less effectively. 
Interestingly, the minimum shifts towards larger $\Delta l$ values for smaller $\Delta k$. This shift occurs because a lower spring constant requires a larger $\Delta l$ to reach the same barrier height; thus, the optimal compromise between pronounced bistability and a moderate barrier (the ``sweet spot'') moves to the right. Concurrently, the width of the dip narrows for smaller $\Delta k$ because the barrier increases more rapidly with $\Delta l$.

Although not explicitly shown in this plot, similar shifting principles apply to the frequency and the attempt rate. The frequency dependence is already evident in Fig.~\ref{fig:penetration_depths}: at low frequencies ($\omega \lesssim 1/\tau_\mathrm{q}$), bistable screening effects are prominent, while at high frequencies the penetration depth converges to the monostable baseline. This occurs because high $\omega$ provides insufficient time for barrier crossing, effectively suppressing state transitions. In this sense, a high driving frequency $\omega$ acts similarly to a low attempt rate $\nu$: in both cases, the barrier is inefficiently crossed, and if $\omega$ is sufficiently large (or $\nu$ sufficiently small), the system reverts to the monostable regime as transitions are kinetically suppressed. The corresponding shift of the optimal $\Delta l$ towards smaller values at higher frequencies can additionally be verified by comparing the $\lambda(\Delta l)$ curves at different $\omega$ (see Fig.~\ref{fig:penetration_vs_Dl_combined} versus Fig.~\ref{fig:penetration_vs_Dl_combined_high_freq}).

The vertical dotted lines in Fig.~\ref{fig:penetration_vs_Dl_combined}(\textbf{a}) indicate the $\Delta l$ values for which the state relaxation time matches the driving frequency, i.e., $\omega \tau_\mathrm{q} = 1$. This ``resonance'' condition serves as an excellent predictor for the position of the minimum, as it marks the point where transitions are still efficiently driven by the external signal. Regarding the model agreement, the LRCT (eq.~\eqref{eq:Penetration_depth}) and the CT (eq.~\eqref{eq:ContinuousBistable}) overlap almost perfectly due to the small force amplitude, which keeps the system well within the linear regime. The BD simulations capture the qualitative trends but exhibit a slight offset, which can be attributed to the weak signal-to-noise ratio inherent to simulations with such small driving forces.

In Fig.~\ref{fig:penetration_vs_Dl_combined}\textbf{b}), we fix $\Delta k = 0$ (states differ only in length) to isolate how an external force assists in overcoming the barrier. The blue curve represents the known small-force limit, matching the LRCT solution. As the force amplitude $F_0$ increases, two trends emerge: First, the minima shift to larger $\Delta l$ values because the external force actively helps to tilt the energy landscape, thereby lowering the effective barrier. Second, the depth of the minimum decreases. Since a single state transition can only absorb a fixed amount of energy (determined by $\Delta l$), the unfolding mechanism becomes increasingly inefficient relative to the large external force. 

Furthermore, large forces push the system out of the linear regime, leading to a clear departure from the LRCT predictions (c.f. Figs.~\ref{fig:bode_plots_large_force} and \ref{fig:penetration_vs_Dl_Dk_large_force}). A crucial aspect of this nonlinearity is the physical constraint of the excitation fraction $q$, which is bounded between $0$ and $1$ and cannot be arbitrarily modified by the external force. Despite this nonlinearity, the CT and the BD simulations remain in excellent agreement, validating our continuous model even beyond the linear response limit.

\subsection{Large-force regime and applicability limits of linear theory}
\label{sec:Large_force_limits}

\begin{figure}[h]
\centering
\includegraphics[width=\columnwidth]{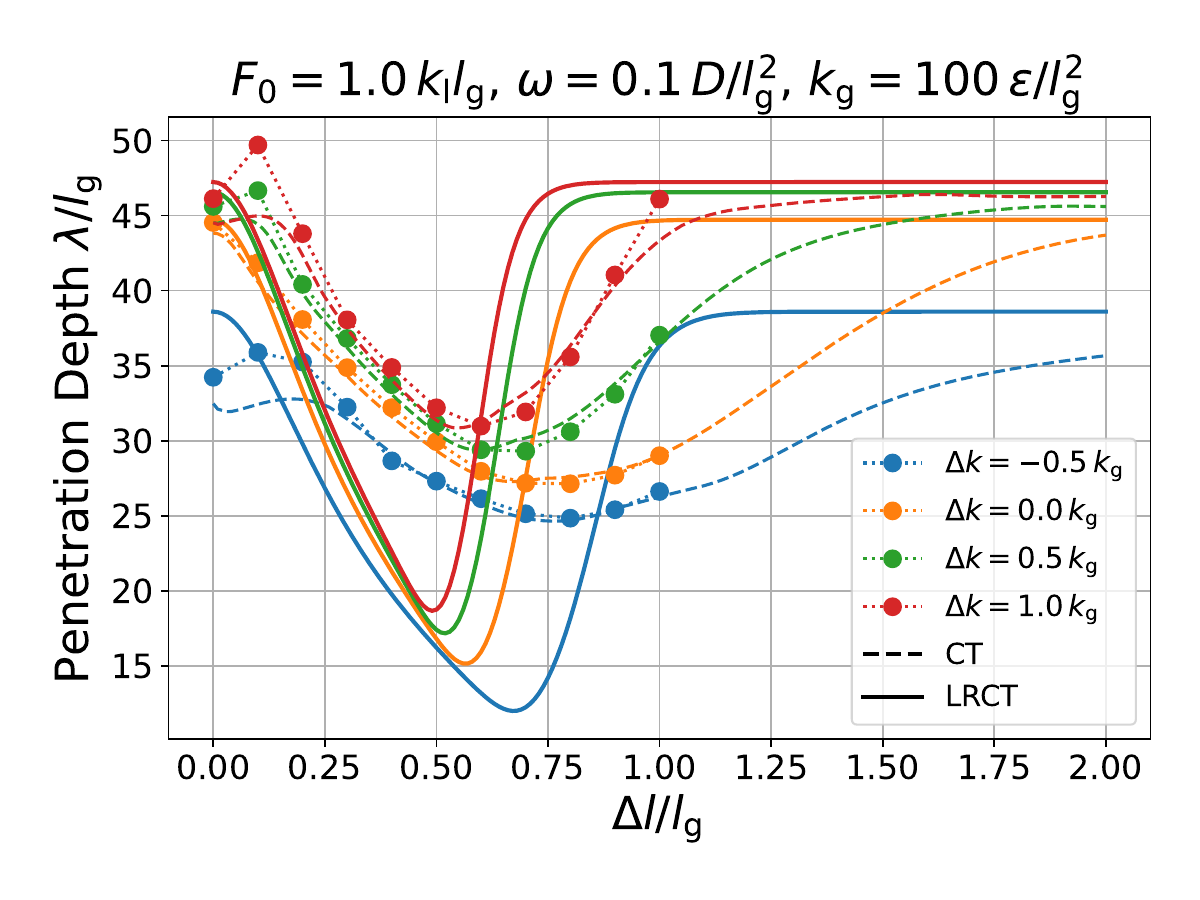}
\caption{
\textbf{Dependence of penetration depth on $\Delta l$ and $\Delta k$ at large force.}
Same as Fig.~\ref{fig:penetration_vs_Dl_combined}(\textbf{a}) but for $F_0 = 1.0\,k_\mathrm{g}l_\mathrm{g}$ and $\nu = 10.0\,D/l_\mathrm{g}^2$.
\label{fig:penetration_vs_Dl_Dk_large_force}
}
\end{figure}

Fig.~\ref{fig:penetration_vs_Dl_Dk_large_force} displays the spatial penetration depth $\lambda$ as a function of conformational length $\Delta l$ for a large driving force ($F_0 = 1.0\,k_\mathrm{g}l_\mathrm{g}$). While LRCT provided excellent agreement in the small-amplitude regime ($F_0 = 0.1\,k_\mathrm{g}l_\mathrm{g}$, Fig.~\ref{fig:penetration_vs_Dl_combined}\textbf{a}), driving the system with a large force pushes it far beyond linear response conditions. Comparing the LRCT predictions with the non-linear models (BD and CT) reveals two striking discrepancies. First, LRCT predicts significantly deeper screening dips than observed in BD and CT. Second, the screening minima in BD and CT are shifted toward noticeably larger values of $\Delta l$. Because the analytical LRCT expression for $\lambda$ [eq.~\eqref{eq:Penetration_depth}] is independent of the driving amplitude $F_0$, the solid LRCT curves in Fig.~\ref{fig:penetration_vs_Dl_Dk_large_force} are identical to those in Fig.~\ref{fig:penetration_vs_Dl_combined}(\textbf{a}). The shift of the BD and CT curves relative to LRCT thus represents a physical trend that emerges systematically with increasing force amplitude.

The fundamental failure of LRCT at large forces stems from neglecting the physical boundaries of the local excitation fraction $q(X,t) \in [0,1]$. LRCT evaluates the linear susceptibility $\chi_q$ near $f_\mathrm{ext} \approx 0$ and extrapolates this high sensitivity linearly to large force amplitudes. Consequently, LRCT overestimates the excitation—predicting unphysical values of $q$ far outside $[0,1]$—which leads to an exaggerated bistable length variation $q\Delta l$ and artificially deep screening dips. In reality, state transitions can absorb only a limited amount of energy. Under strong forcing, local excitations saturate within $[0,1]$, causing the length changes due to unfolding to lose relative significance compared to the strong mechanical stretching of the spring network. Although bistable effects remain present, their relative contribution diminishes, and the system response increasingly resembles that of a monostable network. This mechanism is directly supported by the large-force Bode plots in Appendix Fig.~\ref{fig:bode_plots_large_force}: while LRCT predicts a severely overestimated excitation response $\chi_q$, the actual excitation in BD saturates, causing the mechanical extension $\chi_x$ to approach the monostable reference curve.

In sharp contrast, the full non-linear continuum theory (CT, eqs.~\eqref{eq:ContinuousBistable} and \eqref{eq:dq_new}) inherently respects the bounded state kinetics $q \in [0,1]$ and non-linear transition rates $k_\pm(f)$, thereby maintaining excellent quantitative agreement with microscopic BD simulations. The shift of the screening minima toward larger $\Delta l$ values in BD and CT highlights the interplay between energy capacity and barrier crossing. On one hand, large forces require large conformational changes $\Delta l$ for the additional bistable length change to provide noticeable attenuation against the large external drive. On the other hand, it is precisely the large external force that enables the system to cross these large barriers in the first place, as strong forcing heavily tilts the double-well potential and dramatically accelerates state switching even for high intrinsic energy barriers (large $\Delta l$).

A further crucial non-linear effect omitted by LRCT is the dynamic coupling between excitation and local stiffness via the stiffness contrast $\Delta k = k_\mathrm{e} - k_\mathrm{g}$. As shown in Fig.~\ref{fig:penetration_vs_Dl_combined}(\textbf{a}) for small forces, a positive stiffness contrast ($\Delta k > 0$) stiffens the excited state. Under large forces, local excitation $q(X,t)$ deviates strongly from equilibrium and varies along the chain. Regions with higher excitation become locally stiffer, transmitting stress more efficiently and locally increasing the penetration depth $\lambda$. This creates a spatially dependent spring constant $k_\mathrm{eff}(X,t)$ and, consequently, a spatially dependent complex wavevector $\tilde{k}(X,t)$—a key non-linear effect captured by CT, whereas LRCT relies on a constant wavevector based solely on $k_\mathrm{eff}^\mathrm{eq}$.

Finally, while CT accurately accounts for these non-linear force effects, both continuum descriptions (CT and LRCT) reach their applicability limits at extremely high driving frequencies ($\omega \gg 1/\tau_\mathrm{mech}$). In this regime, the spatial penetration depth drops below the lattice spacing ($\lambda \lesssim l_\mathrm{g}$), violating the continuum assumption of a slowly varying strain field ($\partial_X u \ll 1$). Consequently, discrete lattice effects dominate: although BD and CT both correctly predict vanishing response amplitudes ($\chi_x, \chi_q \to 0$), the phase responses of extension and excitation in discrete BD simulations diverge from continuous field predictions.


\section{Discussion: Design Principles for Adaptive Mechanical Attenuation}
The quantitative agreement of our model allows us to formulate a practical design protocol for materials with frequency-dependent attenuation, highlighting two complementary capabilities that distinguish these systems from classical harmonic oscillators. Our analysis reveals that bistable internal kinetics creates \emph{both} a tunable mechanical skin depth \emph{and} a frequency-independent response window, with both effects controlled by the ratio of mechanical relaxation and switching times. This dual functionality represents the central novelty of our work: while a purely harmonic or monostable overdamped chain exhibits a penetration depth rigidly tied to its baseline stiffness ($\lambda \sim (k_{\mathrm{g}}\omega)^{-1/2}$) and a monotonically decaying response, the bistable chain decouples the attenuation characteristics from static rigidity and introduces a plateau regime where the mechanical compliance remains buffered against frequency changes.

Importantly, because the energy barrier controls the transition rates (via the Arrhenius factor in $r_{\mathrm{ge}}$ and $r_{\mathrm{eg}}$), the kinetic rate $\nu$ can only be chosen after the screening strength—and thus the barrier—has been fixed. Consequently, the design steps must be ordered as follows:

\begin{enumerate}
		\item \textbf{Baseline rigidity:} Define the required mechanical response in the high-frequency limit. For $\omega \gg 1/\tau_{\mathrm{mech}}$ the chain behaves like a passively damped harmonic system, so the ground state spring constant $k_{\mathrm{g}}$ is selected to match the desired stiffness and high-frequency damping properties. This step simultaneously sets the upper boundary of the response plateau, as $\tau_{\mathrm{mech}} \sim N^2 \xi/k_{\mathrm{g}}$ determines where mechanical relaxation completes.
		\item \textbf{Tunable screening intensity:} Set the strength of the attenuation relative to the baseline through mechanical parameters alone. The penetration depth $\lambda$ eq.~\eqref{eq:Penetration_depth} is reduced by a factor
\begin{equation}
r = \frac{\lambda_{\mathrm{bi}}}{\lambda_{\mathrm{mono}}} \approx \frac{1}{1 + \gamma/2},
\end{equation}
where $\gamma = \beta (\Delta l)^2 q_{\mathrm{eq}}(1-q_{\mathrm{eq}}) k_\mathrm{eff}^\mathrm{eq}$ quantifies the coupling to internal degrees of freedom. A desired reduction in signal penetration can thus be engineered by adjusting the magnitude of the conformational change $\Delta l$, the stiffness contrast $\Delta k$, or the inverse temperature $\beta$. However, since the energy barrier rises rapidly (exponentially with $\sim \beta (\Delta l)^2$), excessively large conformational changes can lead to \emph{state-locking}, where transitions are effectively suppressed. In this regime, the attempt frequency $\nu$ required to overcome the barrier becomes physically inaccessible, thereby rendering the kinetic tuning in Step 3 unfeasible.

    \item \textbf{Tunable activation frequency:} Finally, the microscopic rate $\nu$ is chosen to match the target frequency $\omega_{0}$ (via $\omega_{0}\tau_{\mathrm{q}}\approx 1.5$). Since the barrier height is already fixed from Step 2, $\nu$ allows for the precise positioning of both the screening kink and the plateau boundary along the frequency axis without altering the attenuation intensity.
\end{enumerate}

This machinery allows for the design of mechanical materials that retain high transparency to low-frequency vibrations (useful for sensing) while selectively attenuating high-frequency noise and maintaining frequency-insensitive compliance in an intermediate band. The design principles are broadly applicable because the governing parameters span multiple orders of magnitude across different physical realizations. Effective spring constants $k_\mathrm{g}$ range from $\sim 10^{-2}$~pN/nm in soft, pH-responsive DNA origami nanosprings~\cite{Karna_2021} to $\sim 10^3$~pN/nm in stiff molecular links or compressed biopolymers~\cite{Xingfei_2005}, and extend further to macroscopic metamaterials. Notably, even within DNA origami systems, spring constants can vary by orders of magnitude depending on the specific architecture: coiled nanospring designs with i-motif elements yield much softer responses ($\sim 0.03$--$0.5$~pN/nm)~\cite{Karna_2021}, while tile-based reconfiguration mechanisms exhibit significantly stiffer behavior~\cite{Chen_2014,Bae_2014}. The state relaxation time $\tau_\mathrm{q}$—which is the directly observable quantity in experiments, implicitly containing the barrier information and the attempt frequency $\nu$—varies even more dramatically: from nanoseconds in fast-switching polymer segments~\cite{Zhao_2025,Tanaka_2024} to seconds in DNA origami folding~\cite{Bae_2014} and minutes for reconfiguration processes~\cite{Chen_2014}, with additional tunability through temperature~\cite{Pawlak_2019}. Importantly, these timescales can vary significantly even for similar systems depending on geometry, concentration, and environmental conditions, so the values cited here represent order-of-magnitude estimates rather than precise predictions. The mechanical relaxation time $\tau_\mathrm{mech} \sim N^2 \xi / k_\mathrm{g}$ is even more tunable: beyond the variations in friction $\xi$ and stiffness $k_\mathrm{g}$, the quadratic dependence on the chain length $N$ allows shifting $\tau_\mathrm{mech}$ over many decades independently of $\tau_\mathrm{q}$. This vast parameter space ensures that the condition for the response plateau ($\tau_\mathrm{mech} \ll \tau_\mathrm{q}$) and the screening kink ($\omega \tau_\mathrm{q} \approx 1.5$) can be realized in diverse systems, from single-molecule setups and bistable polymer networks to engineered mechanical metamaterials. While complete parameter sets require systematic characterization within single systems, the predicted screening and plateau effects span frequency ranges relevant for soft robotics and mechanosensing applications.

\section{Conclusion}
We have developed and rigorously validated a comprehensive theoretical framework for bistable overdamped spring media with coupled deformation and internal-state dynamics. Starting from microscopic Brownian dynamics with Poisson two-state dynamics, we derived a continuum theory that captures both mechanical wave propagation and kinetic state switching on equal footing.

Our key findings are threefold. First, the theory achieves quantitative agreement with BD simulations across dynamic response functions, linear response spectra, and spatial penetration profiles, establishing its fidelity as a predictive modeling tool. Second, analytical solutions reveal that bistability introduces \emph{two complementary phenomena}: a universal screening mechanism that reduces penetration depth $\lambda$ at frequencies $\omega\tau_\mathrm{q}\lesssim 1$, and a frequency-insensitive response plateau emerging from timescale separation ($\tau_\mathrm{mech} \ll \tau_\mathrm{q}$). While the screening strength is controlled by $\gamma \propto (\Delta l)^2$, the characteristic frequency scale for \emph{both} effects is governed by the state relaxation time $\tau_\mathrm{q}$, which depends directly on the attempt frequency $\nu$. Third, the onset of screening occurs at the universal scale $\omega\tau_\mathrm{q}\approx 1.5$, enabling the attenuation regime and plateau position to be shifted across decades via $\nu$ without altering the material's baseline rigidity.

These results extend the theoretical description of bistable media beyond inhomogeneous domain wall motion \cite{Krumhansl_1975, Wada_1978, Kedia_2023} to a platform for controlled signal attenuation. By bridging the gap between phenomenological hysteresis models \cite{Paulsen_2025a, Muhaxheri_2024} and microscopic kinetics, we establish bistable chains as \emph{tunable media with frequency-dependent damping}. By engineering the bistability strength $\gamma$ and switching kinetics $\nu$, mechanical signals can be screened within a desired frequency band while remaining transparent outside it.

The generic nature of our model---requiring only overdamped dynamics, bistable elements, and diffusive transport---suggests broad applicability beyond the specific chain geometry studied here. We anticipate that similar screening and plateau phenomena should emerge in two- and three-dimensional bistable networks \cite{Roller_2024}, (active) gels with conformational switching \cite{Duarte_2023, Duarte_2024, Skarsetz_2022}, metamaterials, and even granular materials with force-dependent contact states\cite{Liu_2024, Liu_2024b}. This universality, combined with analytical tractability, enables quantitative predictions for diverse experimental platforms without system-specific recalibration.

The presented framework naturally generalizes to multi-stable elements, nonlinear driving regimes, or active fluctuations, providing a rigorous physical basis for constructing soft metamaterials with rationally tailored, frequency-dependent wave propagation. Since the required kinetic timescale separation relies on fundamental barrier-crossing statistics rather than dynamic control loops, experimental realizations using existing platforms---such as DNA origami springs, bistable polymer networks, or colloidal assemblies \cite{Skarsetz_2022, Chen_2018}---appear well within reach. Such physical model systems hold significant potential for applications requiring targeted mechanical filtering and passive frequency-selective damping.

From a broader perspective, this work aligns with the ongoing demand in soft robotics and biomedical engineering for materials that intrinsically couple targeted energy dissipation with their base structural mechanics \cite{Vanderborght_2008, Pratt_1995}. Rather than relying solely on external control algorithms or complex series elastic actuators to ensure safe human-robot interaction \cite{Laffranchi_2009a, Culmer_2010} or to passively exploit natural dynamics \cite{Vanderborght_2006, Jafari_2011}, our findings demonstrate that distinct frequency-filtering characteristics can be directly embedded into the material's micro-kinetics through careful structural design.
For instance, the ability to damp specific frequency bands is analogous to the passive tuning required for tremor suppression in rehabilitation devices \cite{Case_2013, Laffranchi_2013} or vibration isolation in precision machinery \cite{Ast_2009, Sun_1995}. By selecting appropriate kinetic parameters, a soft actuator can be designed to be compliant to low-frequency contact forces (safety) yet stiff to high-frequency disturbances (stability), effectively creating a passive impedance controller embedded within the material matter \cite{Schiavi_2008, Kajikawa_2012}.

\quad\\
{\it Acknowledgments --}
The authors thank Nils G\"oth and Johannes Steppe for inspiring discussions and a critical reading of the manuscript. The authors acknowledge support from the state of Baden-W\"urttemberg through bwHPC, and from the German Research Foundation (DFG) both under the reference `INST 39/1232-1 FUGG' (bwForCluster NEMO 2) and through the Research Unit FOR 5099 Reducing complexity of nonequilibrium systems (project number 431945604).

\appendix
	\bibliographystyle{apsrev4-2-noeprint}
	\bibliography{Referenzen, Literatur}

\counterwithin{figure}{section}
\captionsetup[figure]{justification=justified,singlelinecheck=false}

\captionsetup[figure]{
    format=plain,
    indention=0pt,
    justification=centering,
    singlelinecheck=false
}

\section{Graphical Abstract}
\begin{center}
\includegraphics[width=0.49\textwidth]{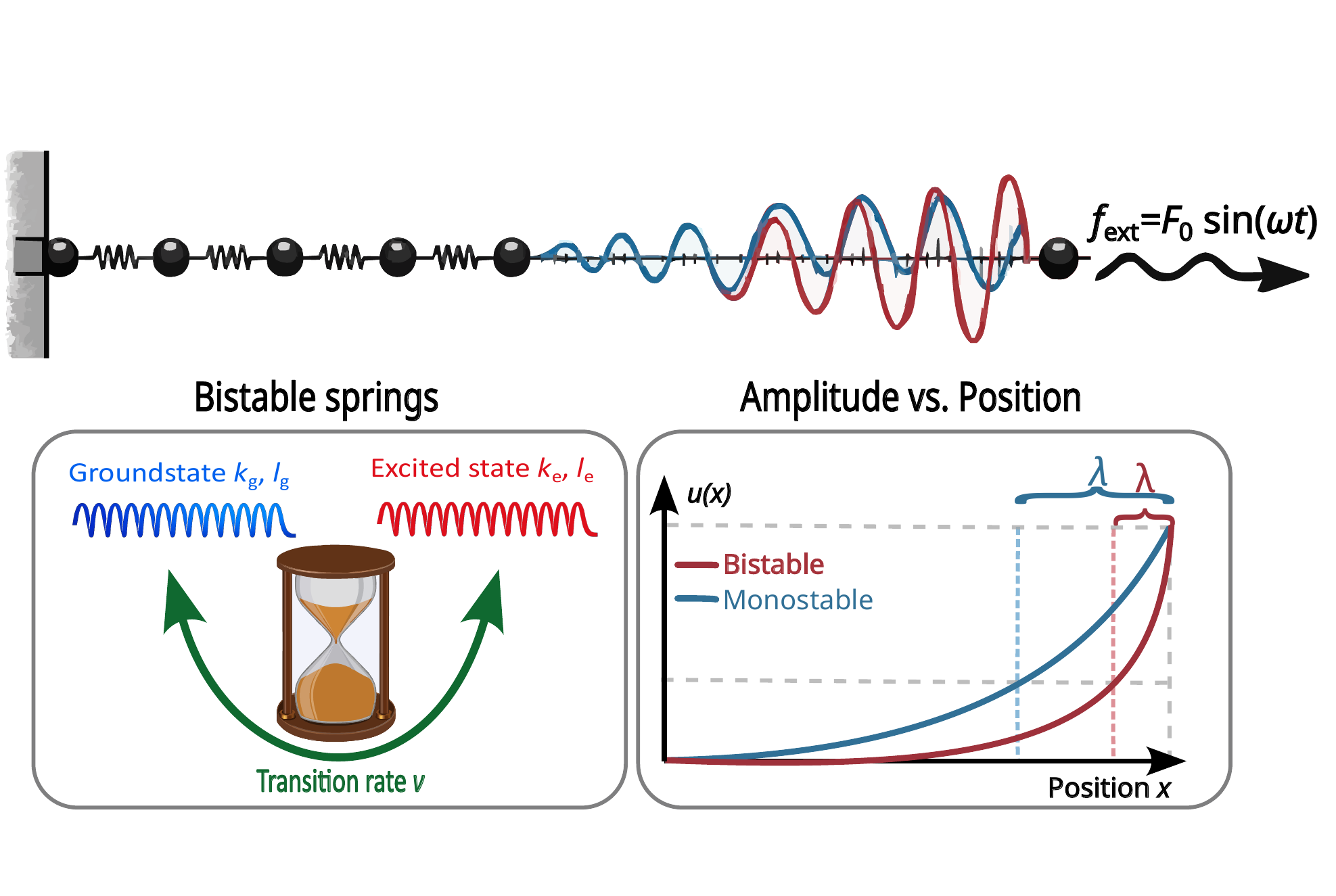}
\captionof{figure}{Graphical Abstract}
\label{fig:graphical_abstract}
\end{center}

\section{Monostable solution for periodic driving}
\label{sec:Monostable}

In the monostable limit, the continuum theory simplifies to the diffusion equation \eqref{eq:ContinuousMonostable}. Here, we derive the analytical solution for a periodic external force (see also \cite{Xu_2023}). The chain is fixed at the left end, $u(0,t) = 0$, and subject to the boundary condition
\begin{equation}
\left. k_\mathrm{g} \partial_X u\left(X,t \right)\right|_{X=L} = F_0\sin\left(\omega t \right)
\end{equation}
at the right end. We assume the chain starts in equilibrium, $u(X,0) = 0$.

A reasonable ansatz that satisfies both boundary conditions is the quasi-static response $u_\mathrm{p}\left(X,t\right) = \frac{F_0}{k_\mathrm{g}} X \sin\left(\omega t \right)$. However, this function does not satisfy the equation of motion. Therefore, we introduce a correction term $v(X,t)$ such that
\begin{equation}
u\left(X,t \right) = u_\mathrm{p}\left(X,t \right) + v\left(X,t \right).
\end{equation}
To preserve the boundary conditions, $v$ must satisfy the homogeneous counterparts $v(0,t) = 0$ and $\left.\partial_X v\right|_{X=L} = 0$. Substituting this ansatz into Eq.~\eqref{eq:ContinuousMonostable} yields an inhomogeneous PDE for the correction term:
\begin{equation}
\dot{v} - \alpha \partial^2_X v = - \dot{u}_\mathrm{p} = -\frac{F_0}{k_\mathrm{g}} X\omega \cos\left(\omega t \right).
\end{equation}

We expand $v$ in terms of the spatial eigenfunctions $\sin(\lambda_n X)$ with time-dependent coefficients $T_n(t)$:
\begin{equation}
v\left(X,t \right) = \sum\limits_{n=1}^\infty T_n\left(t \right) \sin\left(\lambda_n X \right),
\end{equation}
where the eigenvalues $\lambda_n = \frac{(2n-1)\pi}{2L}$ automatically fulfill the homogeneous boundary conditions. Inserting this expansion into the PDE for $v$ and exploiting the orthogonality of the sine functions yields a first-order ordinary differential equation for each mode $n$:
\begin{equation}
	\begin{split}
\dot{T}_n\left(t \right) + \alpha \lambda_n^2 T_n &= \frac{2}{L} \int\limits_{0}^L \left(-\frac{F_0}{k_\mathrm{g}} X\omega \cos\left(\omega t \right) \right) \sin \left(\lambda_n X\right) \mathrm{d}X\\ &= \frac{2F_0\omega}{Lk_\mathrm{g}} \frac{\left(-1 \right)^n}{\lambda_n^2} \cos\left(\omega t \right).
	\end{split}
\end{equation}

The solution to this ODE is the sum of a homogeneous and a particular part:
\begin{equation}
T_n = A_n e^{-\alpha \lambda_n^2 t} + \frac{2F_0\omega\left(-1 \right)^n}{Lk_\mathrm{g}\lambda_n^2} \frac{\omega \sin\left(\omega t \right) + \alpha \lambda_n^2 \cos\left(\omega t \right)}{\left(\alpha \lambda_n^2 \right)^2 + \omega^2}.
\end{equation}

The integration constants $A_n$ are determined by the initial condition $u(X,0) = 0$, which requires $v(X,0) = 0$ and therefore $T_n(0) = 0$. This yields:
\begin{equation}
	\begin{split}
A_n &= - \frac{2F_0\omega\left(-1 \right)^n}{Lk_\mathrm{g}\lambda_n^2} \cdot\frac{\alpha \lambda_n^2}{\left(\alpha \lambda_n^2 \right)^2 + \omega^2}\\
&= \left(-1 \right)^{n+1} \frac{2F_0\omega\alpha}{Lk_\mathrm{g} \left(\left(\alpha \lambda_n^2 \right)^2 + \omega^2\right)}.
	\end{split}
\end{equation}

Combining the particular ansatz and the correction term, this directly leads to the final result presented in the main text as Eq.~\eqref{eq:PeriodicForceFinalSolution}.

\section{Derivation of the barrier-limited transition rates}
\label{sec:Berglund}

\begin{figure}[htbp]
\centering
\includegraphics[width=\columnwidth]{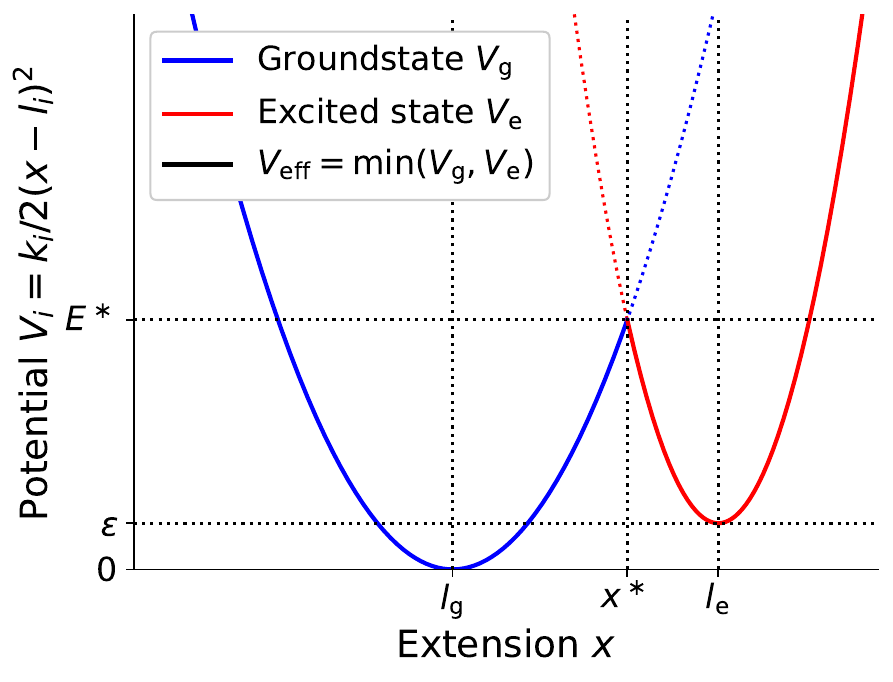}
\caption{
\textbf{Schematic energy landscape of a bistable spring.} 
The potentials of the ground state $V_\mathrm{g}$ and the excited state $V_\mathrm{e}$ are shown for $\Delta l > 0$ and $\Delta k > 0$ in the absence of an external force ($f_\mathrm{ext} = 0$). The effective potential $V_\mathrm{eff} = \min(V_\mathrm{g}, V_\mathrm{e})$ is represented by the solid lines, while the dashed lines indicate the energetically unfavorable branches. The equilibrium extensions $l_\mathrm{g}$ and $l_\mathrm{e}$, as well as the barrier position $x^*$ (the first intersection point of the parabolas), are marked on the extension axis. The vertical axis indicates the reference energy $0$, the intrinsic energy barrier $\epsilon$, and the barrier energy $E^*$.
\label{fig:energy_landscape}
}
\end{figure}

In this section, we derive the barrier-limited transition rates $k_\mathrm{ge,B}$ and $k_\mathrm{eg,B}$ introduced in Eq.~\eqref{eq:Berglund_rates}
Following the framework for non-differentiable potentials by Berglund \cite{Berglund_2009}, the mean first passage time from the ground to the excited state is given by
\begin{equation}
\mathbb{E}_{x_\mathrm{g}}[\tau_\mathrm{e}] = \frac{\int e^{-\beta V(x)} h(x) \, \mathrm{d}x}{\text{cap}(x_\mathrm{g}, x_\mathrm{e})},
\end{equation}
where $h(x)$ is an indicator function for the ground state well, and the denominator denotes the capacity, representing the probability flux over the barrier. The transition rate is the inverse of this expected passage time.

The integral in the numerator corresponds to the occupation probability near the minimum $x_\mathrm{g}$. Using the energies and positions of the minima defined in the main text, this local contribution is obtained via a quadratic expansion around the minimum and the Laplace method:

\begin{equation}
\int e^{-\beta V_i(x)} dx \sim
\sqrt{\frac{2\pi}{\beta k_i}} e^{-\beta E_i},
\quad i \in \{\mathrm{g}, \mathrm{e}\}.
\end{equation}

The calculation of the capacity requires the evaluation of the probability flux across the barrier. As illustrated in Fig.~\ref{fig:energy_landscape}, the effective potential $V_\mathrm{eff}(x) = \min(V_\mathrm{g}(x), V_\mathrm{e}(x))$ exhibits a non-differentiable cusp at the barrier position $x^*$, defined by the intersection of the two individual potential wells, $V_\mathrm{g}(x^*) = V_\mathrm{e}(x^*)$. In the schematic, the energetically favorable branches are represented by solid lines, forming $V_\mathrm{eff}$, while the dashed lines indicate the unfavorable branches. Since the external force term $-fx$ cancels out in the equality $V_\mathrm{g}(x^*) = V_\mathrm{e}(x^*)$, the barrier position $x^*$ is independent of $f$. Depending on the spring constants, the intersection point is given by the following case distinction:

\begin{equation}
x^* =
\begin{cases}
l_\mathrm{g} + \frac{k_\mathrm{e}\Delta l \pm \sqrt{k_\mathrm{g}k_\mathrm{e}\Delta l^2 - 2\epsilon\Delta k}}{\Delta k}, & k_\mathrm{g} \neq k_\mathrm{e} \\[2ex]
l_\mathrm{g} + \frac{\Delta l}{2} + \frac{\epsilon}{2k_\mathrm{g}\Delta l}, & k_\mathrm{g} = k_\mathrm{e}
\end{cases}
\end{equation}

To compute the flux, we expand the potential around the barrier using $z = x - x^*$:
\begin{equation}
\label{eq:V_approx}
\begin{aligned}
V(x) &\approx V^* + a_\mathrm{g} z + \frac{1}{2} k_\mathrm{g} z^2, \quad z < 0, \\
V(x) &\approx V^* + a_\mathrm{e} z + \frac{1}{2} k_\mathrm{e} z^2, \quad z > 0,
\end{aligned}
\end{equation}
with the slopes at the barrier
\begin{equation}
a_\mathrm{g} = k_\mathrm{g}(x^* - l_\mathrm{g}) - f,
\qquad
a_\mathrm{e} = k_\mathrm{e}(x^* - l_\mathrm{e}) - f.
\end{equation}

We calculate the capacity based on the microscopic Metropolis scheme with attempt frequency $\nu$. The flux from the ground to the excited state is defined as
\begin{equation}
\text{cap}(x_\mathrm{g}, x_\mathrm{e}) = \nu \int_{x_\mathrm{g}}^{x_\mathrm{e}} \rho_\mathrm{g}\left(x \right) p_\mathrm{ge}(x) \mathrm{d}x,
\end{equation}
where $\rho_\mathrm{g}\left(x \right)$ is the probability density of a particle being in state $g$ at extension $x$, and $p_\mathrm{ge}(x) = \min \left(1, e^{-\beta\left(V_\mathrm{e}\left(x\right) - V_\mathrm{g}\left(x\right)\right)} \right)$ is the Metropolis acceptance probability. For extensions below the barrier ($x < x^*$), the transition requires crossing an energy difference, so $p_\mathrm{ge} = e^{-\beta\left(V_\mathrm{e} - V_\mathrm{g}\right)}$. Above the barrier ($x \ge x^*$), the transition is always accepted, $p_\mathrm{ge} = 1$.
\begin{equation}
\begin{split}
\text{cap}(x_\mathrm{g}, x_\mathrm{e})
&= \nu \left(\int_{x_\mathrm{g}}^{x^*} e^{-\beta V_\mathrm{e}\left(x \right)} \mathrm{d}x
+ \int_{x^*}^{x_\mathrm{e}}e^{-\beta V_\mathrm{g}\left(x \right)} \mathrm{d}x\right).
\end{split}
\end{equation}
Note that evaluating the capacity in the reverse direction (from $e$ to $g$) yields the exact same result, which ensures the symmetry of the flux. In the respective integration intervals, the potentials do not have a global minimum; instead, their dominant contribution arises from the vicinity of the barrier $x^*$. Applying the piecewise approximations in Eq.~\eqref{eq:V_approx} and the Laplace method, we obtain:
\begin{equation}
\begin{split}
\text{cap}(x_\mathrm{g}, x_\mathrm{e})
&= \frac{\nu}{\beta} e^{-\beta V^*}
\left(\frac{\left|a_\mathrm{g}\right| + \left|a_\mathrm{e}\right|}{\left|a_\mathrm{g}a_\mathrm{e}\right|}\right).
\end{split}
\end{equation}

Finally, the barrier-limited transition rate from the ground to the excited state is given by the capacity divided by the occupation probability of the ground state:
\begin{equation}
\begin{split}
k_\mathrm{ge,B} &= \frac{\text{cap}(x_\mathrm{g}, x_\mathrm{e})}{\sqrt{\frac{2\pi}{\beta k_\mathrm{g}}} e^{-\beta E_\mathrm{g}}}\\
&= \nu \sqrt{\frac{k_\mathrm{g}}{2\pi\beta}} \left(\frac{\left|a_\mathrm{g}\right| + \left|a_\mathrm{e}\right|}{\left|a_\mathrm{g}a_\mathrm{e}\right|}\right) e^{-\beta\left(E^*-E_\mathrm{g} \right)}.
\end{split}
\end{equation}
Analogously, the rate for the reverse transition is:
\begin{equation}
\begin{split}
k_\mathrm{eg,B}
&= \nu \sqrt{\frac{k_\mathrm{e}}{2\pi\beta}} \left(\frac{\left|a_\mathrm{g}\right| + \left|a_\mathrm{e}\right|}{\left|a_\mathrm{g}a_\mathrm{e}\right|}\right) e^{-\beta\left(E^*-E_\mathrm{e} \right)}.
\end{split}
\end{equation}
As discussed in the main text, combining these barrier-limited rates with the diffusion-limited rates ensures detailed balance and correctly captures the transition dynamics for arbitrary barrier heights.

\section{Derivation of Linear Response and Penetration Depth}
\label{app:LinearResponseDerivation}

\subsection{Solution of the Linearized Boundary Value Problem}
\label{app:LinearResponseSolution}

We start from the linearized coupled equations in Fourier space Eq.~\eqref{eq:LinearizedCoupled} in the main text.
The second equation can be solved algebraically for $\tilde{q}$ in terms of the strain field $\partial_X \tilde{u}$:
\begin{equation}
\label{app:eq:q_solution}
\tilde{q} = \frac{\gamma}{\Delta l} \frac{1}{1-i\omega\tau_\mathrm{q}+\gamma} \, l_\mathrm{g} \partial_X \tilde{u}.
\end{equation}
Substituting this expression into the first equation eliminates $\tilde{q}$ and yields the decoupled differential equation for the displacement field alone [Eq.~\eqref{eq:DecoupledU} in the main text]:
\begin{equation}
\partial_X^2 \tilde{u} = - \frac{i\omega \xi}{k_\mathrm{eff}^\mathrm{eq} l_\mathrm{g}^2} \left(1+\frac{\gamma}{1-i\omega\tau_\mathrm{q}} \right) \tilde{u}.
\end{equation}
We introduce the complex wave vector $\tilde{k}$ via Eq.~\eqref{eq:ktilde_def}, giving the general solution
\begin{equation}
\tilde{u}(X) = C_1 \sin(\tilde{k} X) + C_2 \cos(\tilde{k} X).
\end{equation}
The boundary condition at the fixed left end, $\tilde{u}(X=0)=0$, implies $C_2 = 0$. The linearized boundary condition at the right end $X = L = N l_\mathrm{g}$ reads
\begin{equation}
\left.k_\mathrm{eff}^\mathrm{eq} \left( l_\mathrm{g} \partial_X \tilde{u} - \tilde{q} \Delta l \right)\right|_{X=L} = \tilde{f}_\mathrm{ext}.
\end{equation}
Using Eq.~\eqref{app:eq:q_solution} again to eliminate $\tilde{q}$, the boundary condition becomes
\begin{equation}
\left.k_\mathrm{eff}^\mathrm{eq} l_\mathrm{g} \left(1-\frac{\gamma}{1-i\omega\tau_\mathrm{q}+\gamma} \right) \partial_X \tilde{u} \right|_{X=L} = \tilde{f}_\mathrm{ext}.
\end{equation}
With $\partial_X \tilde{u} = C_1 \tilde{k} \cos(\tilde{k} X)$, we determine $C_1$ and obtain the displacement at the right end:
\begin{equation}
\tilde{u}(L) = \frac{\tilde{f}_\mathrm{ext}}{k_\mathrm{eff}^\mathrm{eq} l_\mathrm{g} \tilde{k}} \left(1+\frac{\gamma}{1-i\omega\tau_\mathrm{q}}\right) \tan\left(\tilde{k} L \right).
\end{equation}
The linear response function is therefore
\begin{equation}
\chi_x(\omega) = \frac{\tilde{u}(L)}{\tilde{f}_\mathrm{ext}} = \frac{1}{k_\mathrm{eff}^\mathrm{eq} l_\mathrm{g} \tilde{k}} \left(1+\frac{\gamma}{1-i\omega\tau_\mathrm{q}}\right) \tan\left(\tilde{k} N l_\mathrm{g} \right).
\end{equation}

\subsection{Perturbative Expansion of the Penetration Depth}
\label{app:ScreeningDerivation}

The penetration depth is defined as $\lambda = 1/\mathfrak{Im}\tilde{k}$. To obtain an analytical expression, we expand Eq.~\eqref{eq:ktilde_def} for weak bistability ($\gamma \ll 1$). Writing $\tilde{k}^2 = A(1 + \gamma B)$ with
\begin{equation}
A = \frac{i\omega \xi}{k_\mathrm{eff}^\mathrm{eq} l_\mathrm{g}^2}, \qquad B = \frac{1}{1-i\omega\tau_\mathrm{q}},
\end{equation}
we use the Taylor expansion $\sqrt{1+\epsilon} \approx 1 + \epsilon/2$ for small $\epsilon$:
\begin{equation}
\tilde{k} = \sqrt{A} \sqrt{1 + \gamma B} \approx \sqrt{A} \left(1 + \frac{\gamma}{2} B\right).
\end{equation}
The imaginary part is extracted by noting that $\sqrt{i} = (1+i)/\sqrt{2}$:
\begin{equation}
\mathfrak{Im}\tilde{k} \approx \sqrt{\frac{\omega\xi}{2 k_\mathrm{eff}^\mathrm{eq} l_\mathrm{g}^2}} \, \mathfrak{Im}\left[ (1+i) \left(1 + \frac{\gamma}{2} \frac{1}{1-i\omega\tau_\mathrm{q}}\right) \right].
\end{equation}
Evaluating the real part yields
\begin{equation}
\mathfrak{Im}\tilde{k} \approx \sqrt{\frac{\omega\xi}{2 k_\mathrm{eff}^\mathrm{eq} l_\mathrm{g}^2}} \left(1 + \frac{1}{2} \frac{\gamma\left(1+\omega\tau_\mathrm{q} \right)}{1 + \omega^2\tau_\mathrm{q}^2}\right),
\end{equation}
and thus the penetration depth [Eq.~\eqref{eq:Penetration_depth} in the main text]:
\begin{equation}
\lambda = \frac{1}{\mathfrak{Im}\tilde{k}} \approx \sqrt{\frac{2 k_\mathrm{eff}^\mathrm{eq} l_\mathrm{g}^2}{\omega\xi}} \left(1 + \frac{1}{2} \frac{\gamma\left(1+\omega\tau_\mathrm{q} \right)}{1 + \omega^2\tau_\mathrm{q}^2}\right)^{-1}.
\end{equation}

\section{Bode plot for large force}
\begin{minipage}{\columnwidth}
\centering
\includegraphics[width=0.98\textwidth]{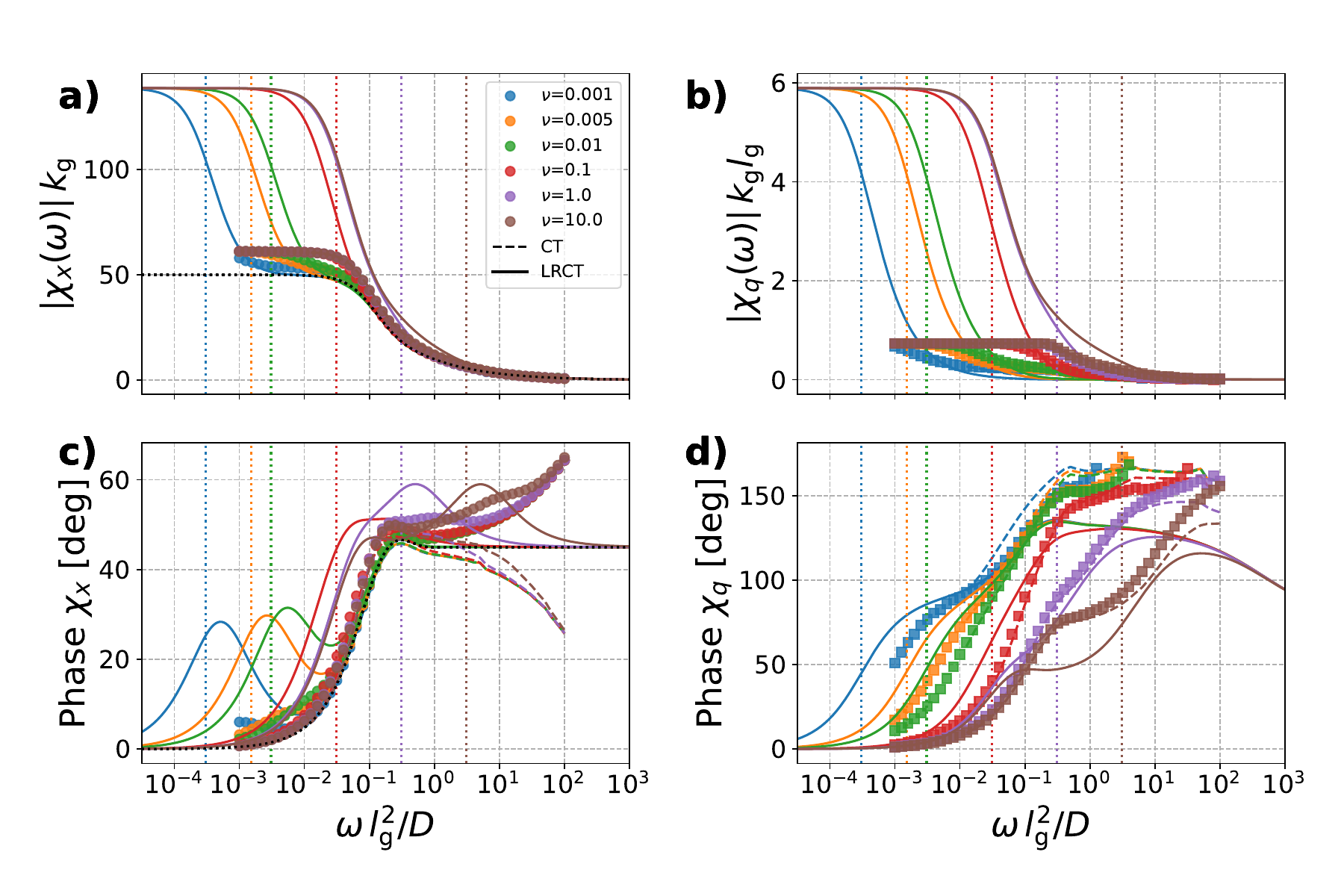}
\captionof{figure}{
\textbf{Bode plots for large driving force.}
Same as Fig.~\ref{fig:bode_plots} but for $F_0 = 1.0\,k_\mathrm{g}l_\mathrm{g}$.
At large forces, the excitation $q$ saturates within $[0,1]$, rendering bistable effects insignificant and causing the system to approach a harmonic oscillator.
Consequently, LRCT (eq.~\eqref{eq:LinearResponse}) fails by overestimating the response due to extrapolating the high sensitivity near $f\approx0$.
In contrast, the full continuum theory (CT eq.~\eqref{eq:ContinuousBistable} and \eqref{eq:dq_new}, crosses) maintains excellent agreement with BD simulations.
}
\label{fig:bode_plots_large_force}
\end{minipage}

\section{Turning points of the penetration depth}
\label{sec:TurningPoints}
We define the transition points (turning points) of the $\lambda(\omega)$ curve in a log-log plot as the zeros of the second derivative $\mathrm{d}^2\log\lambda / \mathrm{d}(\log\omega)^2 \stackrel{!}{=} 0$. For small $\gamma$, the penetration depth can be approximated using $1/(1+x) \approx e^{-x}$:
\begin{equation}
\lambda \approx \sqrt{\frac{2 k_\mathrm{eff}^\mathrm{eq}}{\omega\xi}} \exp\left( -\frac{\gamma/2\left(1+\omega\tau_\mathrm{q} \right)}{1 + \omega^2\tau_\mathrm{q}^2} \right).
\end{equation}
This expression is equal to Eq.~\eqref{eq:Penetration_depth} up to linear order in $\gamma$ and simplifies the logarithmic derivatives tremendously. Taking the first derivative with respect to $\log\omega$ yields the monostable slope of $-1/2$ plus a bistable correction:
\begin{equation}
\frac{\mathrm{d}\log\lambda}{\mathrm{d}\log\omega} = - \frac{1}{2} - \omega\tau_\mathrm{q} \frac{\gamma}{2} \frac{1-\omega^2\tau_\mathrm{q}^2 - 2\omega\tau_\mathrm{q}}{\left(1 + \omega^2\tau_\mathrm{q}^2\right)^2}.
\end{equation}
The second derivative is given by
\begin{equation}
\frac{\mathrm{d}^2\log\lambda}{\mathrm{d}\log\omega^2} = -\omega \tau_\mathrm{q} \frac{\gamma}{2} \frac{\mathrm{d}}{\mathrm{d}\omega} \left[ \omega \frac{1-\omega^2\tau_\mathrm{q}^2 - 2\omega\tau_\mathrm{q}}{\left(1 + \omega^2\tau_\mathrm{q}^2\right)^2} \right].
\end{equation}
Evaluating the derivative and setting the result to zero leads to a fourth-order polynomial in $\omega\tau_\mathrm{q}$, which factors as:
\begin{equation}
\left(\omega^2\tau_\mathrm{q}^2 + (2+2\sqrt{2})\omega\tau_\mathrm{q}-1 \right)\left(\omega^2\tau_\mathrm{q}^2 + (2-2\sqrt{2})\omega\tau_\mathrm{q}-1 \right) = 0.
\end{equation}
Of the four roots, only the positive real solution is physically meaningful for the transition to the screening regime:
\begin{equation}
\omega\tau_\mathrm{q} = -1 + \sqrt{2} + \sqrt{4-2\sqrt{2}} \approx 1.50.
\end{equation}
This indicates that the characteristic dip in the penetration depth occurs when the driving frequency is roughly $1.5$ times the internal relaxation rate $1/\tau_\mathrm{q}$.

\subsection{Additional Plots}
\begin{minipage}{\columnwidth}
\centering
\includegraphics[width=0.98\textwidth]{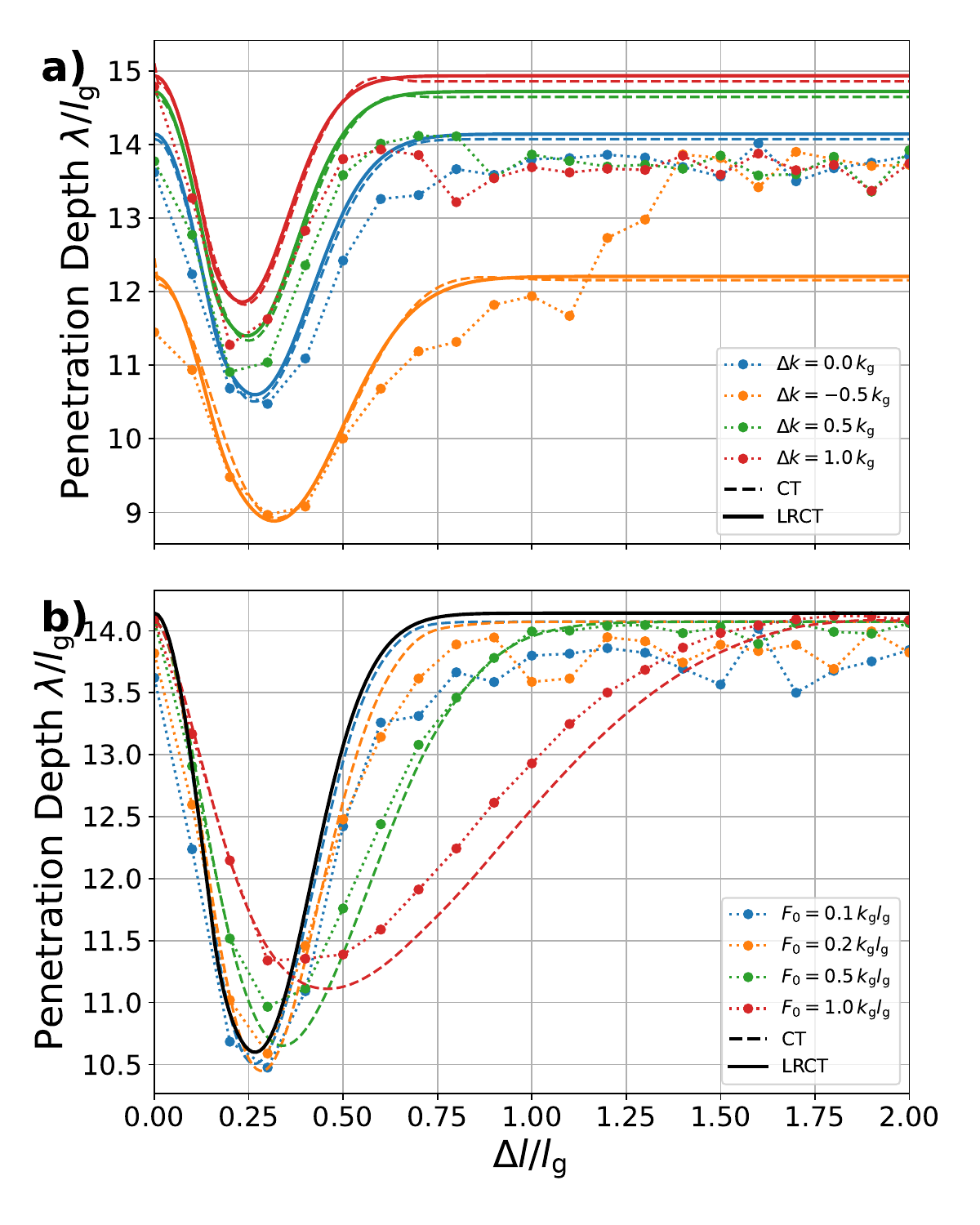}
\captionof{figure}{
\textbf{Dependence of penetration depth on $\Delta l$ at higher frequency.}
Same as Fig.~\ref{fig:penetration_vs_Dl_combined} but for $\omega = 1.0\,D/l_\mathrm{g}^2$.
Vertical dotted lines are omitted because $\omega \tau_\mathrm{q} = 1$ is already satisfied at $\Delta l = 0$ for this frequency.
At higher frequencies, the penetration depth decreases overall and the minima shift to smaller $\Delta l$ values because less time is available for barrier crossing.
Consequently, the relative dip depth (ratio of monostable baseline to minimum) decreases.
\label{fig:penetration_vs_Dl_combined_high_freq}
}
\end{minipage}

\end{document}